\documentclass[pre, showpacs, twocolumn, letterpaper]{revtex4}

\usepackage{graphicx,epsfig}
\usepackage{latexsym}
\usepackage{amssymb, amsmath}

\def\bea{\begin{eqnarray}}
\def\eea{\end{eqnarray}}
\def\be{\begin{equation}}
\def\ee{\end{equation}}

\begin{document}
\title{Function reconstruction as a classical moment problem: A maximum entropy approach} 

\author{Parthapratim Biswas} 
\affiliation{Department of Physics and Astronomy, The University of Southern Mississippi, 
Hattiesburg, MS 39406, USA} 
\email{Partha.Biswas@usm.edu}

\author{Arun K. Bhattacharya} 
\affiliation{Department of Physics, The University of Burdwan, Burdwan, WB 713104, India}


\begin{abstract}
We present a systematic study of the reconstruction of a non-negative function via maximum entropy approach 
utilizing the information contained in a finite number of moments of the function. 
For testing the efficacy of the approach, we reconstruct a set of functions 
using an iterative entropy optimization scheme, and study the convergence profile as the number of 
moments is increased. We consider a wide variety of functions that include a distribution with 
a sharp discontinuity, a rapidly oscillatory function, a distribution with singularities, and 
finally a distribution with several spikes and fine structure. The last example is important 
in the context of the determination of the natural density of the logistic map. 
The convergence of the method is studied by comparing the moments of the approximated functions 
with the exact ones. Furthermore, by varying the number of moments and iterations, we examine to 
what extent the features of the functions, such as the divergence behavior at singular points 
within the interval, is reproduced. The proximity of the reconstructed maximum entropy solution 
to the exact solution is examined via Kullback-Leibler divergence and variation measures for 
different number of moments. 
\end{abstract}

\maketitle

\section{introduction}

The reconstruction of a non-negative distribution from its moments constitutes the so-called 
classical moment problem, and is an archetypal example of an inverse problem~\cite{Shohat, Akheizer} 
in mathematical sciences. Owing to its importance in the context of probability theory and the 
challenging problems of analysis associated with it, the moment problem has attracted the attention 
of a large number of researchers from many diverse fields of science and 
engineering~\cite{Kapur, Lent, Wheeler, Isenberg, Bricogne, Gilmore, Mead, Smith, Drabold, Carlsson, Gotovac}. 
In the classical Hausdorff moment problem (HMP), one addresses the problem of reconstructing 
a non-negative, real valued function $f(x)$ in a finite interval $[a,b]$ from a sequence of 
real numbers. The sequence forms a `moment sequence' that satisfies the Hausdorff 
conditions~\cite{Hausdorff}. The problem is severely 
ill-posed in the Hadamard sense~\cite{Hadamard}. For a finite number of moments, most of the existing numerical 
methods are susceptible to large instabilities but several methods do exist that attempt to construct 
a regularized solution by avoiding these instabilities~\cite{Turek, Mead}. The HMP has been addressed by using a variety 
of methods (such as Tikhonov's regularization 
method~\cite{Tikhonov} and the use of Pollaczek polynomial by Viano~\cite{Vino1, Vino2}) 
but the information-theoretic approach is particularly 
fascinating to the physicists. The latter is based on the \textit {maximum entropy principle} (MEP) 
proposed by Jaynes~\cite{Jaynes}.  The MEP provides a suitable framework to reconstruct a distribution 
by maximizing the Shannon information entropy~\cite{Shannon} and at the same time ensures the matching 
of the moments of the distribution. 
Our interest in the moment problem stems from the fact that inverse problems of this type are 
frequently encountered in many areas of physical, mathematical and biological 
sciences~\cite{Wheeler, Isenberg, Bricogne, Gilmore, Mead, Smith, Schoenfeldt, Steeb, Drabold, Carlsson}. 
A very simple but elegant example is the inversion of the specific heat data of solids.  It 
is known that the constant volume vibrational specific heat of a solid can be expressed as 
the convolution of a known function of the frequency and the vibrational frequency distribution 
function (FDF)~\cite{Maradudin}. The task of extracting the FDF by inverting the experimentally-measured 
values of the specific heat at constant volume as a function of temperature is a well-known example of 
an inverse problem in solid state physics~\cite{Houston, Domb}. 

The focus of our present work is to reconstruct a non-negative function very accurately 
within the framework of the MEP from the knowledge of a finite number of moments. Although there 
exists a number of numerical procedures that address this problem, most of them become unreliable 
when the number of moment constraints exceeds a problem-dependent upper limit. A close review of 
the methods and the study of the example functions presented therein immediately 
reveal the weakness of the methods~\cite{Mead, Turek}. For example, it is very difficult to 
reproduce accurately the van Hove singularities in the frequency distribution of (crystalline) 
solids or the presence of a gap in the density of electronic states in a solid. While the algorithm 
proposed by Silver and R\"{o}der~\cite{Silver} does reproduce the latter correctly and is capable of 
dealing with a large number of moments, we are not aware of any systematic study of function
reconstruction by this approach at this time. It is, therefore, worthwhile to 
explore the possibility of developing a reliable scheme for the entropy optimization program and 
to apply it to a range of non-negative functions having complex structure within the interval. 

The rest of the paper is organized as follows. In Section II we briefly describe a procedure that 
has been developed recently by us to reduce the moment problem to a discretized entropy optimization 
problem (EOP)~\cite{Biswas}.  We then test our methodology in Section III by examining to what extent 
it is successful in reconstructing a wide variety of functions on the basis of input information 
in the form of Chebyshev moments of the functions. The convergence behavior of the maximum entropy 
solution is then discussed in Section IV with particular emphasis on the number of moments. The 
proximity of the reconstructed solution to the exact solution for different distributions is 
also studied via Kullback-Leibler~\cite{Kullback1} divergence and variation measures~\cite{Tag}.

\section{Maximum Entropy approach to the Hausdorff moment problem}

The classical moment problem for a finite interval [a, b], also known as the Hausdorff moment 
problem~\cite{Hausdorff}, can be stated as follows. Consider a set of moments
\be
\mu_i = \int_a^b x^i \, \rho(x)\, dx \quad \quad i = 0, 1, 2, \ldots, m, \quad i \le m
\label{eq-1} 
\ee
\noindent 
of a function $\rho(x)$ integrable over the interval with $\mu_i < \infty $  $\forall \, x \in $ [a,b]. 
The problem is to construct the non-negative function $\rho(x)$ from the knowledge 
of its moments.  The necessary and sufficient conditions for a solution to exist were given by Hausdorff~\cite{Hausdorff}. 
The moment problem and its variants have been studied extensively in the 
literature~\cite{Shohat, Akheizer, Vino1, Schoenfeldt, Wimp}.  Mead and Papanicolaou~\cite{Mead} have, in 
particular discussed a number of moment problems encountered in various of physics.  For a finite number of 
moments, the problem is underdetermined and it is not possible to construct the unique solution from the moment 
sequence unless further assumptions about the function are made. Within the framework of maximum entropy principle, one 
attempts to find a function $\rho(x)$ that maximizes the information entropy functional, 

\be 
S[\rho] = - \int_a^b \rho(x) \,\ln[\rho(x)] \, dx 
\label{eq-2} 
\ee

\noindent 
subject to the moment constraints defined by Eq.~(\ref{eq-1}). The resulting solution is an approximate 
function $\rho_{ME}(x)$, which can be obtained by functional differentiation of a Lagrangian with 
respect to the unknown function $\rho(x)$. The Lagrangian is given by, 
 \be 
L(\rho, \lambda) = -S[\rho] + \sum_{i=0}^m \lambda_i\, \left( \int_a^b x^n \, \rho(x)\, dx - \mu_i \right). 
\label{lag}
\ee 
Now, 
\be 
\frac{\delta L}{\delta \rho(x)} = 0 \: \Longrightarrow \: \rho_{ME}(x) = \exp\left(-\sum_{i=0}^m \lambda_i \, x^i\right). 
\label{eq-3} 
\ee

The normalized function $\rho (x)$ is often referred to as probability density since it is positive 
semidefinite and the interval [a, b] can be mapped onto [0,1] without any loss of generality. 
For a normalized function with $\mu_0$ = 1, the 
Lagrange multiplier $\lambda_0$ is connected to the others via, 
\[
e^{\lambda_0} = \int_0^1 \exp\left(-\sum_{i=1}^m \lambda_i x^i \right) = Z, 
\] 
and the maximum entropy (ME) solution can be written as, 
\be
\rho_{ME}(x) = \exp\left(-\sum_{i=1}^m \lambda_i \, x^i\right)/Z, 
\ee 
where $Z$ is known as the partition function.

A reliable scheme for handling the entropy optimization problem subject to the matching of 
the moments was discussed by us in Ref.\,\cite{Biswas}. The essential idea behind the approach 
is to use a discretized form of the Shannon entropy functional and the moment constraints 
using an accurate quadrature formula. The constraint optimization problem involving 
the primal variables is then reduced to an unconstrained convex optimization program 
involving the dual variables of the problem. This guarantees the existence of a unique solution 
within the framework of maximum entropy principle. The solution is 
{\it least biased}~\cite{note1} and satisfies the moment constraints defined by Eq.\,(\ref{eq-1}). The 
procedure consists of: 1) rewriting the Lagrangian of the problem in Eq.\,(\ref{lag}) in 
terms of the discretized variables to obtain the ME solution, 2) using the 
resulting ME solution in association with Eq.\,(\ref{eq-1}) to reduce the EOP 
as an unconstrained convex optimization problem in dual variables, and finally 3) minimizing 
the objective function in the dual space to obtain the optimal solution in the primal space. 

Using a suitable quadrature (e.g.\,Gaussian) with a set of weights $\omega_j$'s and abscissae $x_j$'s, 
the discretized Lagrangian can be written as, 

\be 
L(\tilde \rho, {\bf \tilde \lambda}) = \sum_{j=1}^n \tilde \rho_j\, \ln \left(\frac{\tilde \rho_j}{\omega_j}\right) - \sum_{i=1}^m \tilde \lambda_i \left(\sum_{j=1}^n t_{ij} \, \tilde \rho_j - \mu_i\right), 
\label{eq-7} 
\ee 
where $ 0 \le \tilde \rho \in R^n$ and $ \tilde \lambda = -\lambda  \in R^m$, respectively are the primal and the dual variables 
of the EOP.  In the equation above, we have used the notation $\tilde \rho_j = \omega_j \rho_j$ and $t_{ij} = (x_j)^i$. 
The discretized ME solution is given by the functional variation with respect to the unknown function as before, 
\be 
\rho^{ME}_j = \exp\left(\sum_{i=1}^m t_{ij} \, \tilde \lambda_i - 1\right), \quad j=1, 2, \ldots n. 
\label{eq-8} 
\ee 

Equations (\ref{eq-1}) and (\ref{eq-8}) can be combined together and the EOP can be reduced to an unconstrained 
convex optimization problem involving the dual variables $\tilde \lambda$'s: 
\be 
\min_{\tilde \lambda \in R^m} \left[ D(\tilde \lambda) \equiv \sum_{j=1}^n \omega_j\,  \exp \left(\sum_{i=1}^m t_{ij} \, \tilde \lambda_i - 1\right) - \sum_{i=1}^m \mu_i \, \tilde \lambda_i \right].  
\ee 

By iteratively obtaining an estimate of $\tilde \lambda$,  $D({\tilde \lambda})$ can be minimized, and the ME 
solution ${\tilde \rho}(\tilde \lambda^{*})$ can be constructed from Eq.~(\ref{eq-8}).  The objective function 
$D(\tilde \lambda)$ can be minimized by modifying a method, which is largely due to Bergman~\cite{Berg}, and 
was presented and discussed at length in Ref.\,\cite{Biswas} both for the power and the Chebyshev moments. 
For the latter, the ME solution can be shown to be expressed in the form of Eq.(\ref{eq-8}) 
with $t_{ij} = T^{*}_i(x_j)$, where $T^*_i(x)$ is the shifted Chebyshev polynomials. In the following, we apply our algorithm 
to reconstruct a variety of functions corresponding to different number of shifted Chebyshev moments.

\section{Application to function reconstruction}
We now illustrate the method by reconstructing a number of exact functions from a knowledge of 
their moments. For all but one of the examples studied here, the moments of the functions can be 
obtained from analytical expressions. In the remaining case the moments have been calculated 
numerically using standard double precision arithmetic.  As mentioned earlier, we map the functions 
onto the interval [0,1] and assume they are normalized so that the functions can be treated as probability 
density functions (pdf) without any loss of generality.  It is well-known that for a finite number 
of moments, the Hausdorff moment problem cannot be solved uniquely.  One needs to supply additional 
information to choose a suitable solution from an ensemble of solutions that satisfy the given moment 
constraints. The maximum entropy (ME) ansatz constructs the {\it least biased} 
solution that maximizes the entropy associated with the density and is consistent with the given moments.  
The accuracy of the reconstructed solution can be measured by varying the number of moments. A comparison with the 
exact solution (if available) would reveal to what extent the ME solution matches with the exact 
solution. For an unknown function with a finite set of moments, the quality of the ME solution may be judged 
by the proximity of the input (exact) moments to the output (approximated) moments resulting 
from the reconstructed distribution. By increasing the number of moments one can systematically 
improve the quality of the solution. It should, however, be noted, that for a function with a complicated 
structure, the convergence of the first few moments does not guarantee its accurate reproduction.  
The ME solution in this case may not represent the exact solution, but is still correct as far as the 
maximum entropy principle is concerned. It is therefore important to study the convergence behavior of the 
solutions with moments for a number of functions with widely different shapes. To this end we 
compare, in the following, our maximum entropy solution corresponding to a variety of exact distribution 
and a distribution amenable to an accurate numerical analysis. 

\subsection{Case 1 : $f(x)$  = 1 }

We begin with a step function which is unity throughout the interval $[0,1]$.  As mentioned earlier, we 
use the shifted Chebyshev polynomials $T^{*}_n(x)$, which is defined via, 

\bea 
T_n^{*}(x) &=& T_n(2x - 1)  \nonumber \\ 
T_n(x) &=& \cos \left[n\cos^{-1}(x))\right] \nonumber \quad \mbox{for} \quad n = 0, 1, \ldots
\eea 

The moments can be calculated analytically in this case, and are given by, 
\[ 
\mu_0 = 1; \quad \mu_1 = 0;  \quad \mu_n = \frac{1 + (-1)^n}{2 - 2n^2} \quad n \neq 1.  
\]

\begin{figure}[htpb]
\includegraphics[width=4.5in, height=3.8in, angle =0]{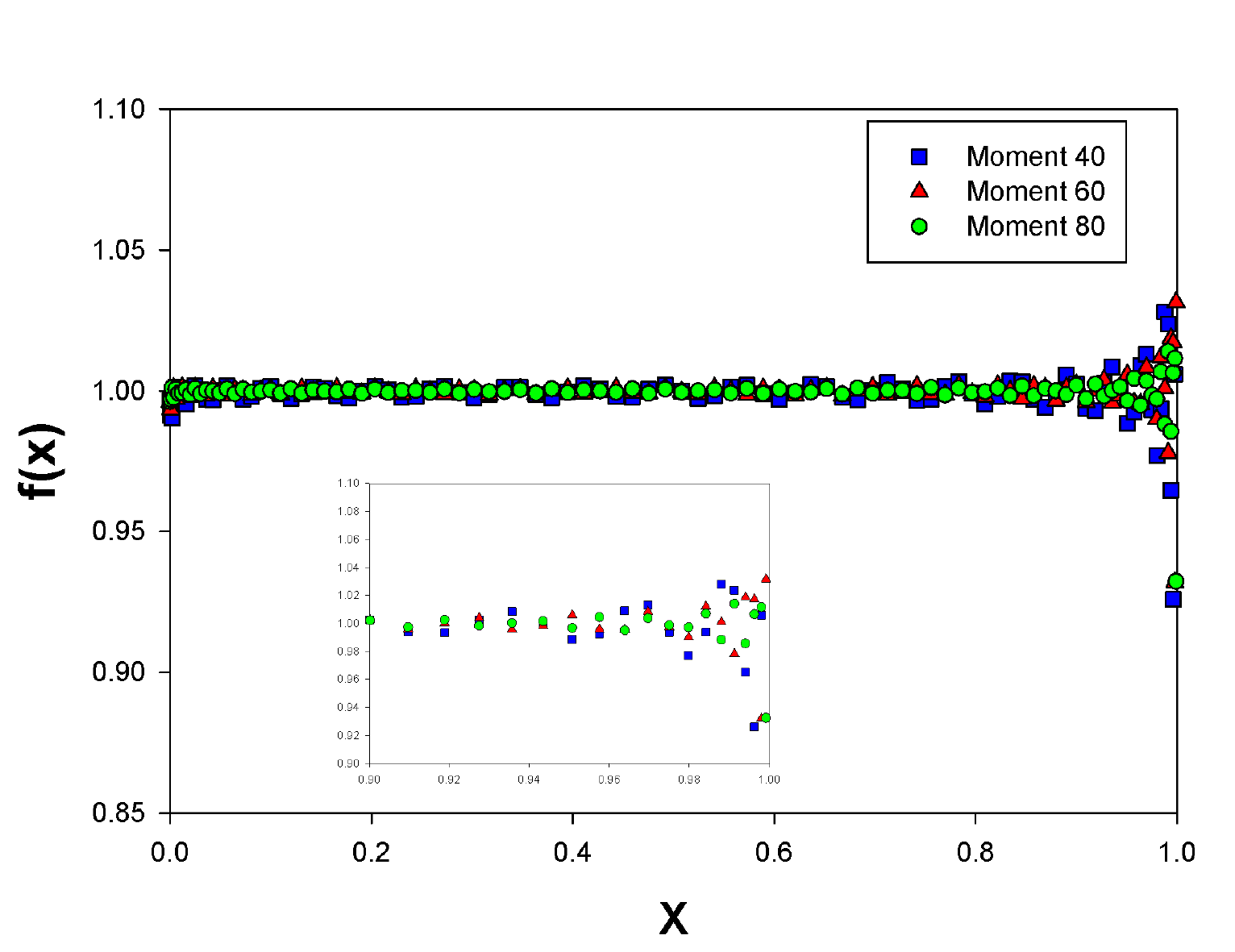}
\caption{
\label{fig1}
(Color online) The step function, $f(x)=1$, reconstructed using the first 40, 60, and 80 Chebyshev moments. 
The plot shows the presence of oscillations at the right edge with decaying amplitude as the number 
of moments increases. The data for 40, 60 and 80 moments are indicated in the figure by boxes (blue), triangles 
(red) and circles (green) respectively. A magnified view of the right edge is shown in 
the inset. 
}
\end{figure}

\begin{figure}[htpb]
\includegraphics[width=4.5in, height=3.8in, angle =0]{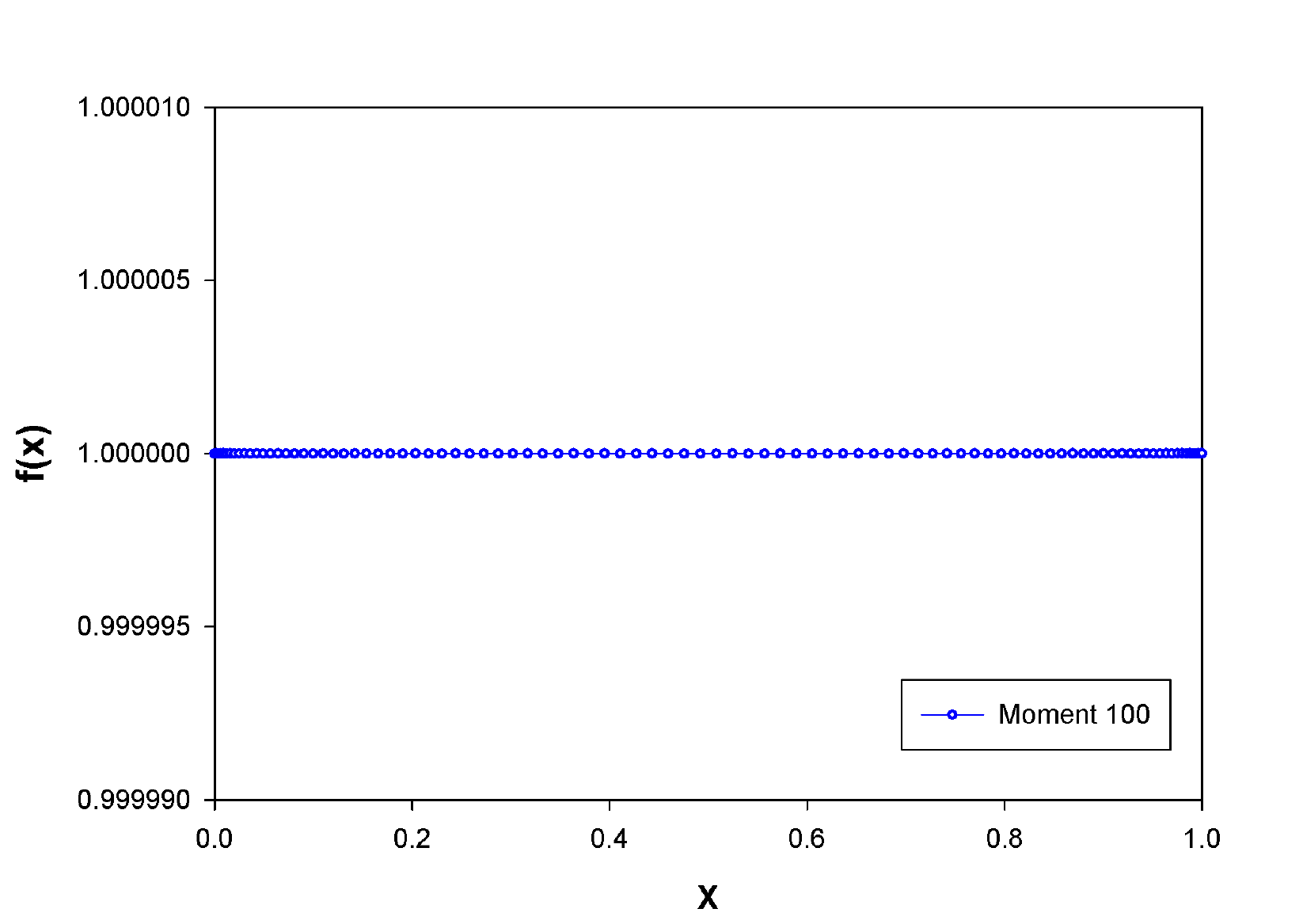}
\caption{
\label{fig2}
(Color online)
The step function as in fig.\ref{fig1} using the first 100 Chebyshev moments. The 
oscillations now completely disappear and the function is reproduced with an error 
less than 1 part in $10^6$.}
\end{figure}

Although the function does not have any structure, it is particularly important because 
of its behavior at the end points.  Owing to the presence of discontinuities at $x=0$ and 1, the 
function is difficult to reproduce close to these points. The sharp discontinuities 
cause the reconstructed function (from a small number of moments) to exhibit spurious 
oscillations near the end points. The oscillations are progressively suppressed by increasing 
the number of Chebyshev moments in our iterative method. Beyond 100 moments the oscillations 
completely disappear. This behavior is seen clearly in fig.\ref{fig1} where we have plotted 
the reconstructed functions corresponding to 40, 60 and 80 moments. The oscillations are 
particularly pronounced as one approaches $x = 1$, but die down with increase in the number 
of moments. The result corresponding to 100 moments is presented in fig.\ref{fig2}. The plot clearly reveals 
that the function has been reproduced with an error, which is less than 1 part in $10^{6}$.  

\subsection{ Case 2 : $f(x) = \frac{3}{2} x^{\frac{1}{2}}$ }

The next example we consider is a square-root function $f(x)=\frac{3}{2}\,x^{\frac{1}{2}}$, where 
the prefactor is chosen to normalize the function. In many physical problems, we often encounter 
distributions showing a square-root behavior. For example, the spectral distribution of a free 
electron gas in 3-dimension is related to the energy via $\sqrt{E}$, and the square-root 
behavior persists in the weak interaction limit (at low energy). It is therefore important 
to see if such a square-root function can be reproduced with a high degree of accuracy using 
our maximum entropy ansatz. The shifted Chebyshev moments for the present case are given by,

\[ 
\mu_n =\frac{9-12 n^2}{9-40n^2 + 16n^4} \quad \mbox{for} \quad n \ge 0. 
\]

\begin{figure}[htpb]
\includegraphics[width=4.5in, height=3.8in, angle=0]{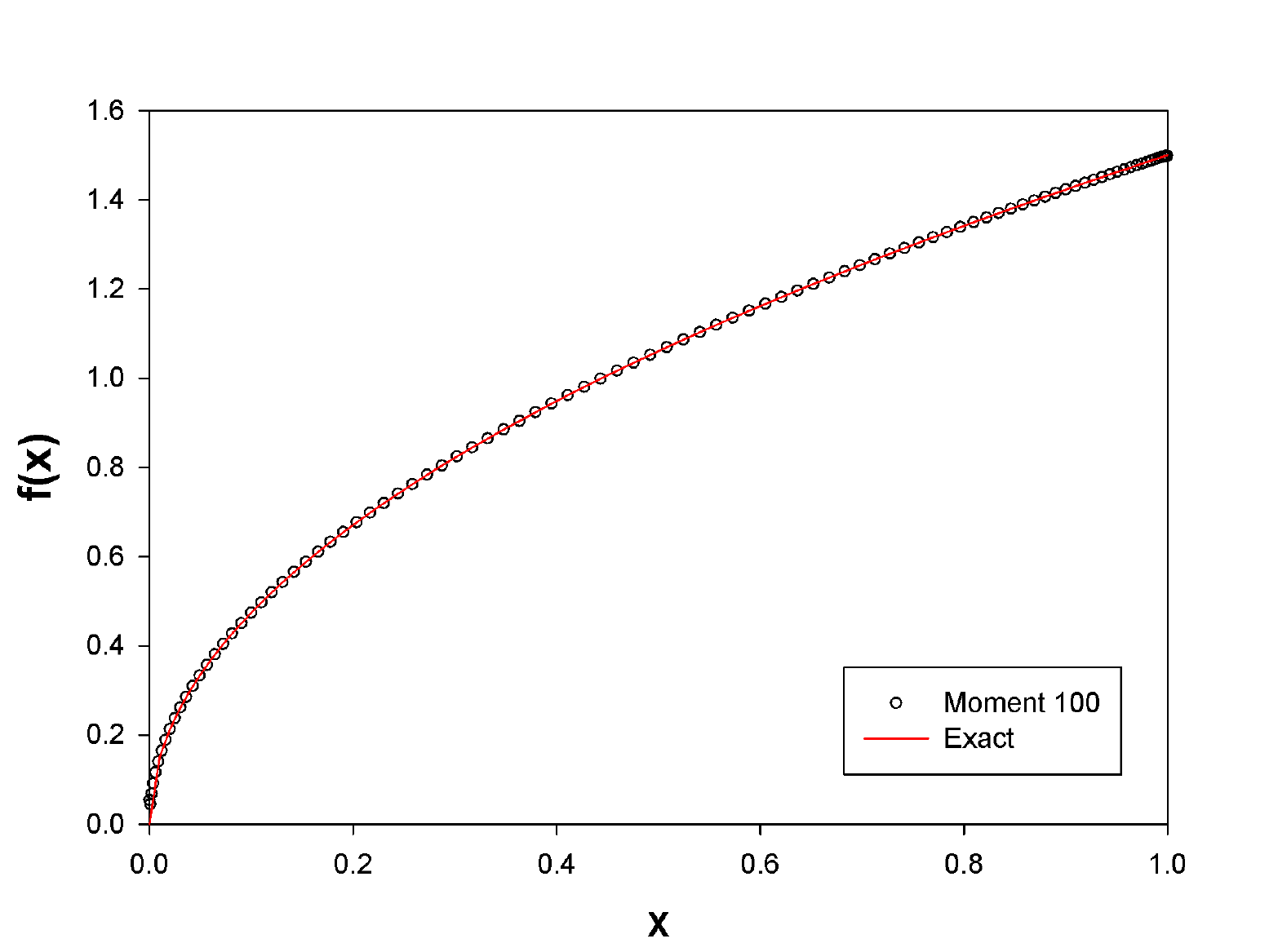}
\caption{
\label{fig3}
The function, $f(x)=\frac{3}{2}x^{\frac{1}{2}}$, and its maximum entropy reconstruction 
using the first 100 Chebyshev moments. For clarity, every second data point is plotted in 
the figure. The exact function is evaluated at the quadrature points and is drawn as 
a line for comparison. 
}
\end{figure}

\begin{figure}[htpb]
\includegraphics[width=4.5in, height=3.8in, angle=0]{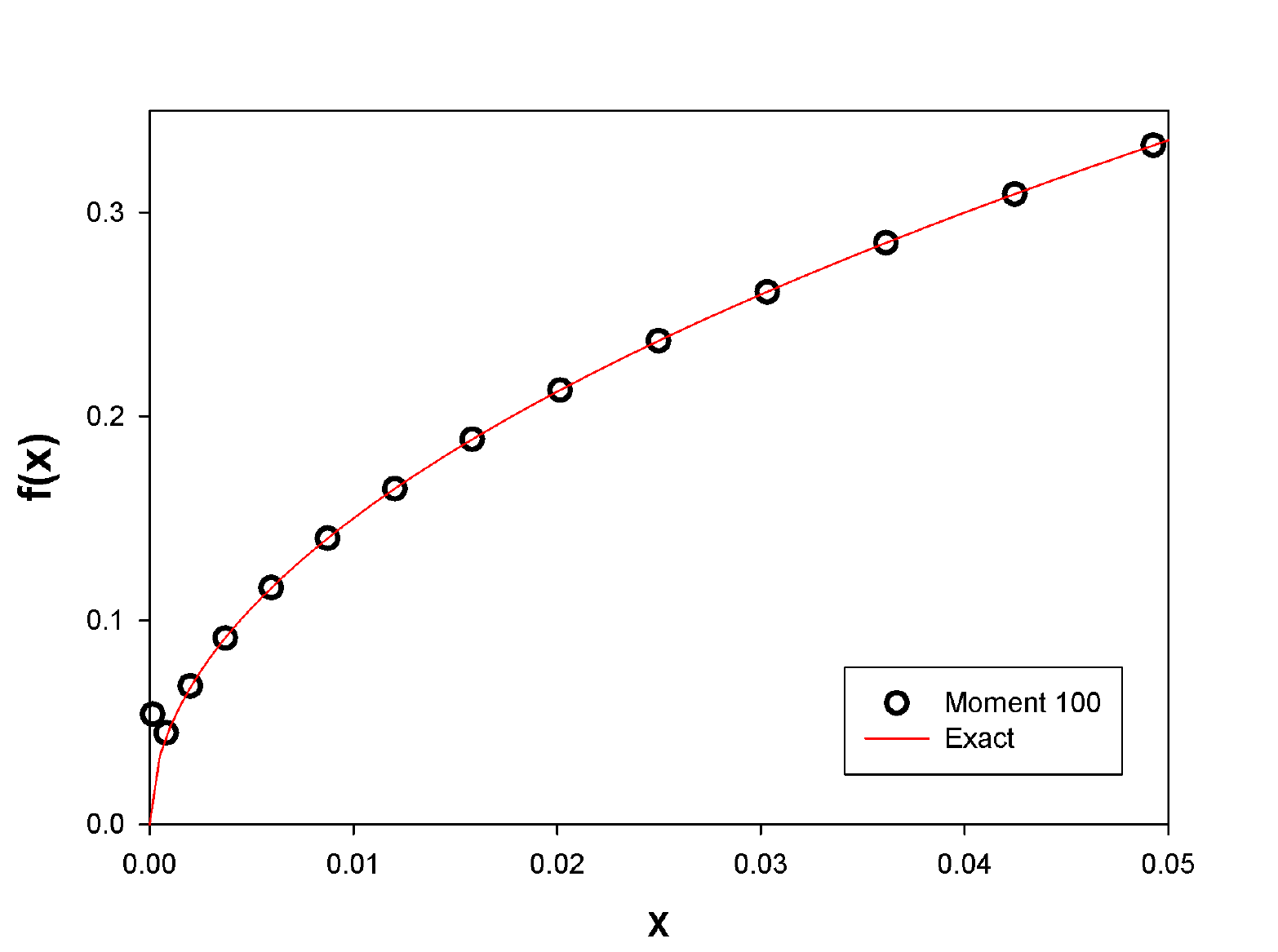}
\caption{
\label{fig4}
The behavior of the function $f(x)=\frac{3}{2}x^{\frac{1}{2}}$  for very small 
values of x along with the exact functional values evaluated at the quadrature points. 
Note that only one point is off the graph indicating an excellent match to the exact 
function for 100 moments. 
}
\end{figure}

\begin{figure}[htpb]
\includegraphics[width=4.5in, height=3.8in, angle=0]{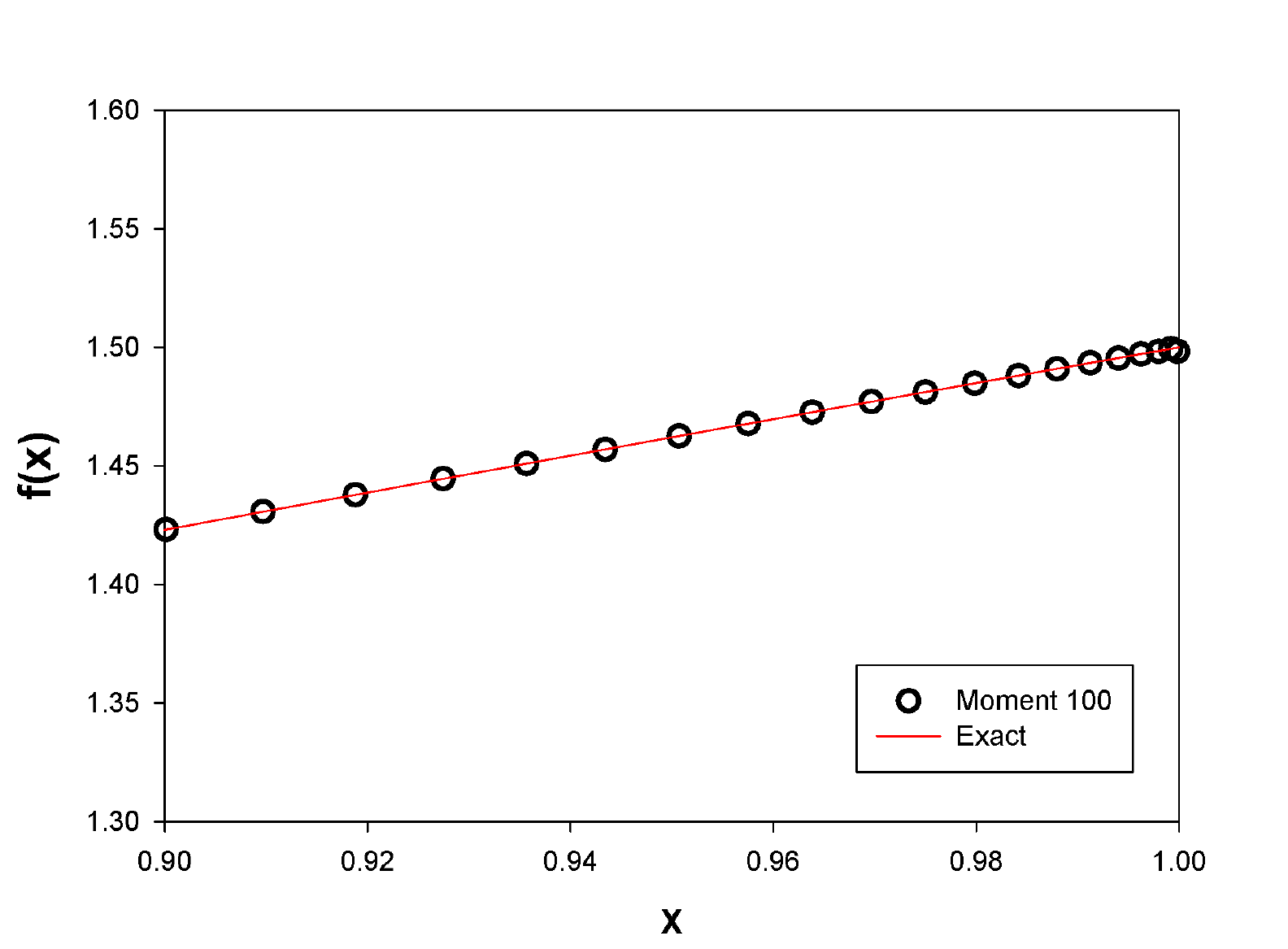}
\caption{
\label{fig5}
The behavior of the function $f(x)=\frac{3}{2}x^{\frac{1}{2}}$ near $x = 1$. 
The exact function is also plotted for comparison.  
}
\end{figure}

The results for the function are plotted in figs. \ref{fig3} to \ref{fig5} for 100 moments.  
The reconstructed function is found to match excellently with the exact function throughout the 
interval as shown in fig.\ref{fig3}. Of particular importance is the behavior of the function near 
$x=0$ and 1.  The square-root behavior is accurately reproduced without any deviation or oscillation
near $x$ = 0 as is evident from fig.\ref{fig4}. Similarly, the behavior near $x$ = 1 is also 
reproduced with a high degree of accuracy as shown in fig.\ref{fig5}. Since our method can exploit 
the information embedded in the higher moments, it is capable of reproducing the function very accurately 
without any oscillation. 

\subsection{Case 3: A double-parabola with a gap}

\begin{figure}[htpb]
\includegraphics[width=4.5in, height=3.8in, angle=0]{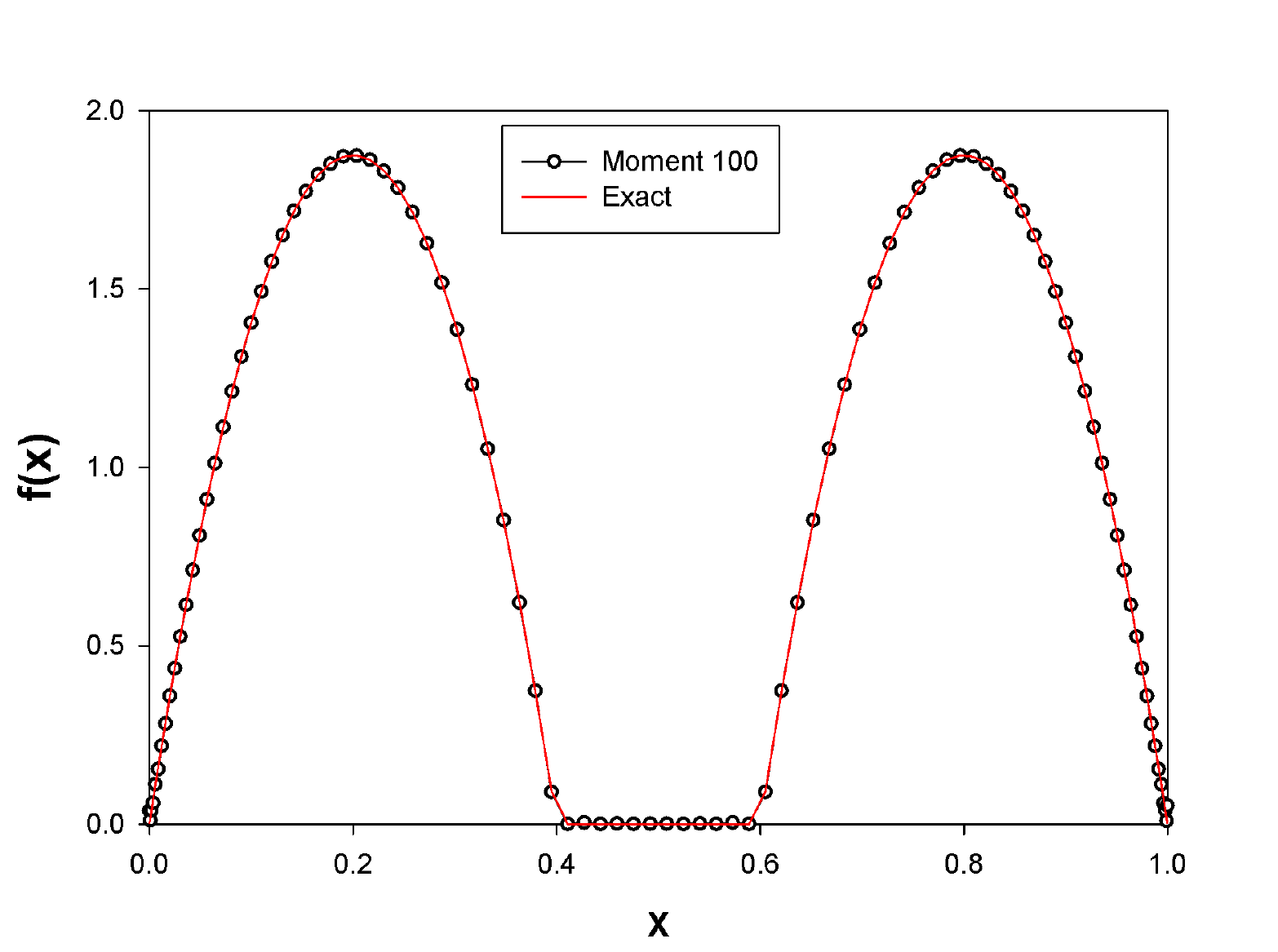}
\caption{
\label{fig6}
The reconstruction of a function with a gap in the interval. The double-parabola 
with a gap is reconstructed using the first 100 moments.  The exact values are 
also plotted in the figure for comparison. 
}
\end{figure}

\begin{figure}[htpb]
\includegraphics[width=4.5in, height=3.8in, angle=0]{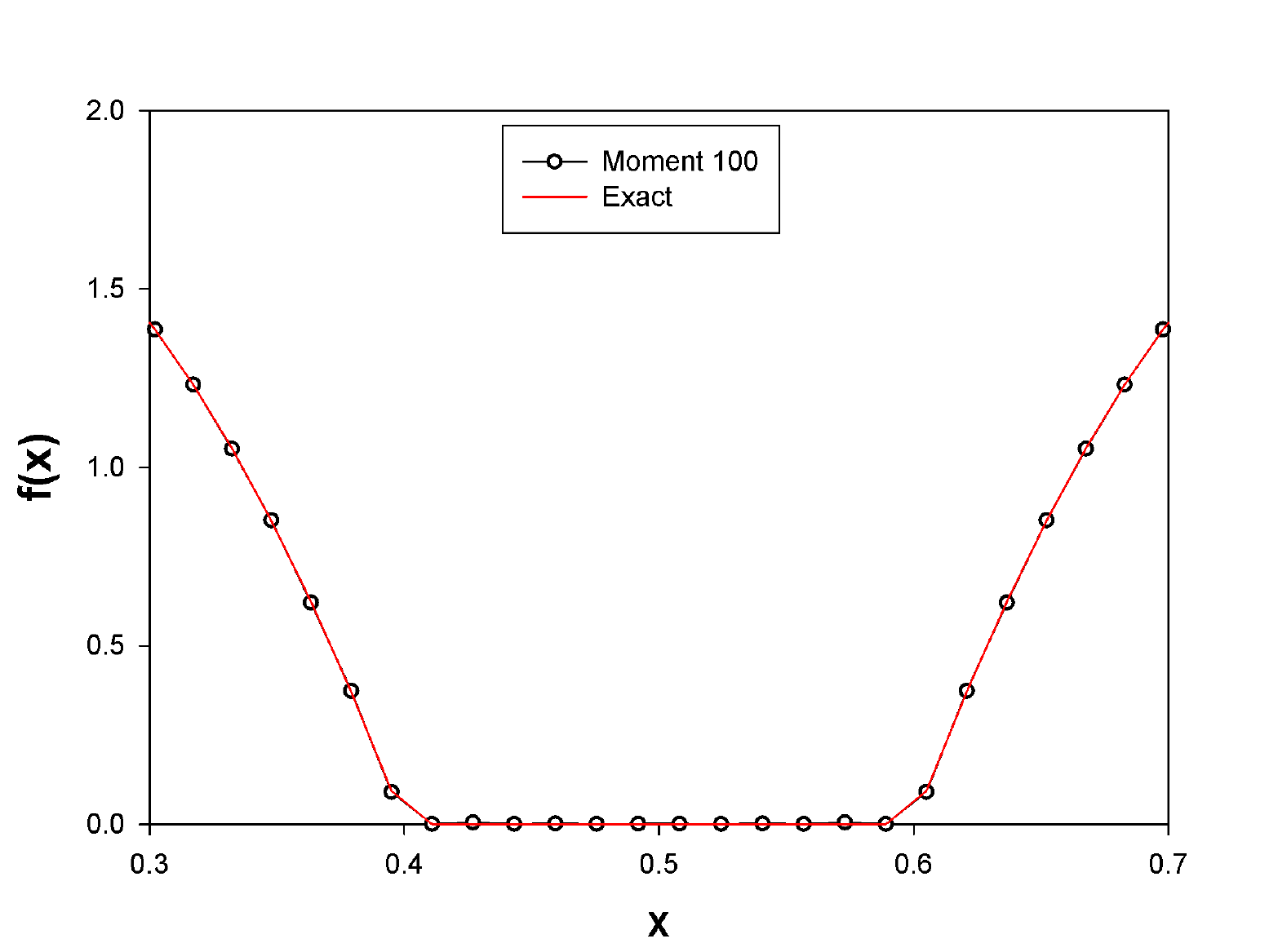}
\caption{
\label{fig7}
The reconstructed double-parabola near the gap along with the exact function 
at the quadrature points. Owing to the finite number of quadrature points, the 
reconstructed function has a non-zero value at $x_1$ = 0.4 and $x_2$ = 0.6.  } 
\end{figure}

Having discussed two relatively simple examples, we now consider a case where the function vanishes in a finite 
domain within the interval. Such a function appears frequently in the context of the energy density of states 
of amorphous and crystalline semiconductors. It is instructive to study whether our maximum entropy (ME) method 
is capable of reproducing a gap in the energy eigenvalue spectrum. Since the moments of the electronic density of states can be obtained from 
the Hamiltonian of the system, our method can be used as an alternative tool to construct the density of states 
from the moments.  This is particularly useful for treating a large non-crystalline system (e.g. in the amorphous or 
liquid state), in which case the direct diagonalization of the Hamiltonian matrix is computationally 
overkill and scales with the cubic power of the system size. In contrast, our maximum entropy ansatz provides 
an efficient and accurate procedure for the determination of total (band) energy and the Fermi level subject 
to the availability of the energy moments. Here we use a toy model of a density of states that consists of two parabolae 
separated by a gap to illustrate the usefulness of our method.  In particular, we choose a normalized distribution with a gap 
from $x_1$ to $x_2$, 

\begin{equation*} 
f(x) = 
\begin{cases}   
A \, x\, (x_{1} - x) & \text{for $x \le x_1$} \\ \\
B \, (x-x_{2})\, (1-x) & \text{for $x \ge x_2$,}
\end{cases}
\end{equation*}

where A and B are given by
\[ 
A=\frac{6}{x^{2}_{1}(1 + x_1 - x_2)}; \quad B = \frac{6}{(1-x_{2})^{2}(1+x_1 -x_2)}.
\] 
In the present case, we choose $x_{1}=\frac{2}{5}$ and $x_{2}=\frac{3}{5}$ giving the value of the gap $(x_2 - x_1) = \frac{1}{5}$.  
The Chebyshev moments of the function can be calculated exactly, and as in the previous examples 
the function is reconstructed from the moments. In figs.~\ref{fig6} and \ref{fig7} we 
have plotted the results obtained from our ME ansatz along with the exact functional values at 
the quadrature points. It is remarkable to note that the reconstructed function matches 
excellently with the exact one. Furthermore, the method reproduces the gap between the 
parabolae correctly without any oscillation in the gap. Table 1 lists the size of the gap 
corresponding to different number of moments for two sets of Gaussian points. Since we are 
using a finite number of quadrature points, the accuracy of our gap size is limited by the 
resolution of the (non-uniform) quadrature grid near the gap. We have chosen a tolerance 
$\epsilon = 5.0 \times 10^{-3}$ for the reconstructed functional value to locate the onset of the gap 
(i.e. the zero of the function)~\cite{note2}.  It is evident from table 1 that as the number of 
moments increases, the size of the gap improves and eventually converges very close to the 
exact numerical value.  The accuracy can be improved further by using more Gaussian points in 
the quadrature. 

\begin{center}
\begin{table}[htbp]
\caption{\label{tab2} Numerical values of the gap for different number of moments from the 
reconstructed double-parabolic distribution.}
\begin{ruledtabular}
\begin{tabular}{lllcc} Moments & 96 points &  192 points \\
\hline
20 & 0.1622 & 0.1676\\
40 & 0.1813 & 0.1875\\
60 & 0.1821 & 0.1902\\
80 & 0.1823 & 0.1909\\
100 &  -- & 0.1922\\
\hline
Exact numerical & 0.1941 & 0.1945\\
\end{tabular}
\end{ruledtabular}
\end{table}
\end{center}

\subsection{Case 4: $ f(x) = \frac{1}{\pi \sqrt{(x - x^2)}}$} 

\begin{figure}[htpb]
\includegraphics[width=4.5in, height=3.8in, angle = 0]{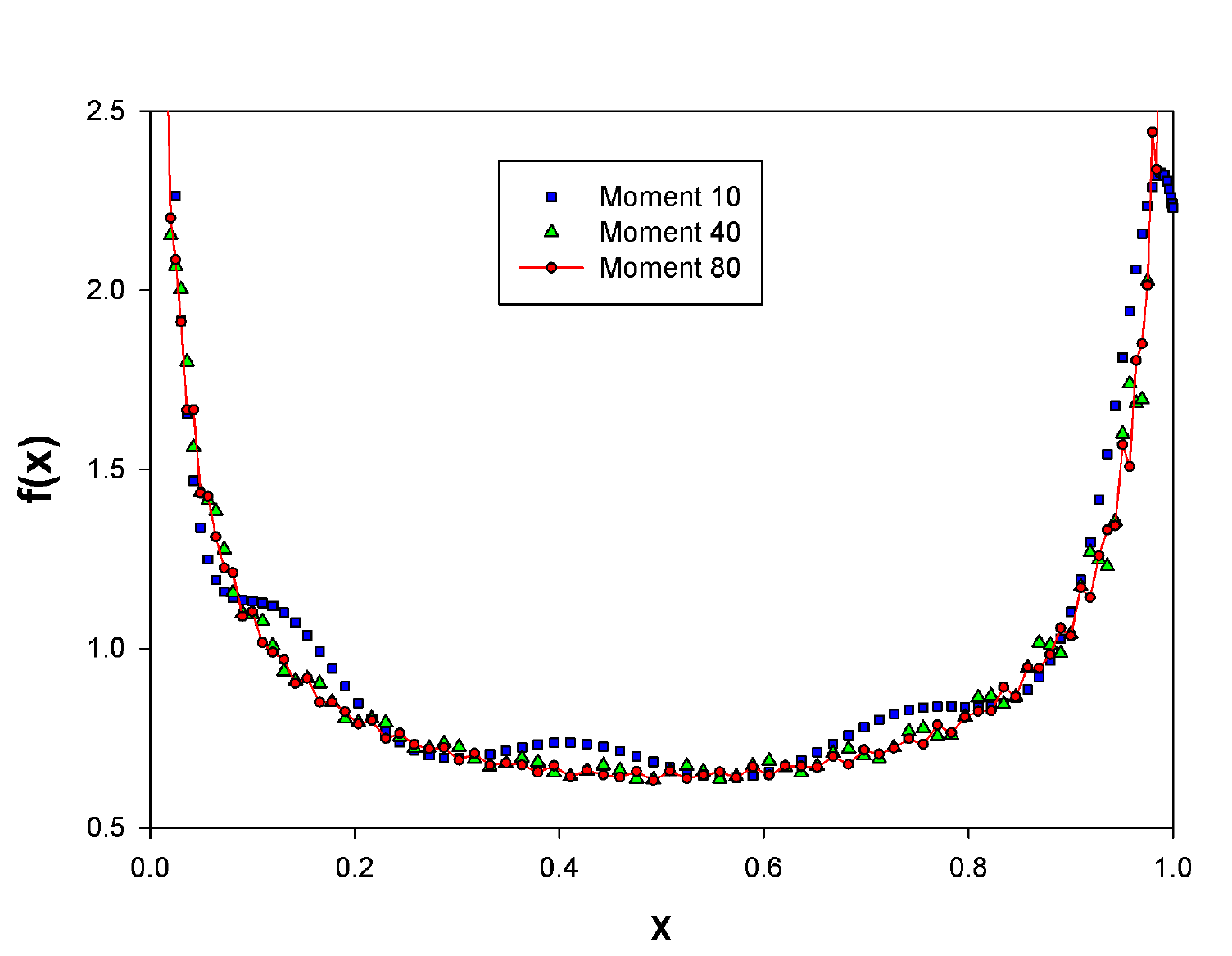}
\caption{
\label{fig8}
(Color online) Reconstruction of the U-function as defined in the text. The data correspond to the 
first 10, 40 and 80 moments as indicated in the plot.
}

\end{figure}

\begin{figure}[htpb]
\includegraphics[width=4.5in, height=3.8in, angle = 0]{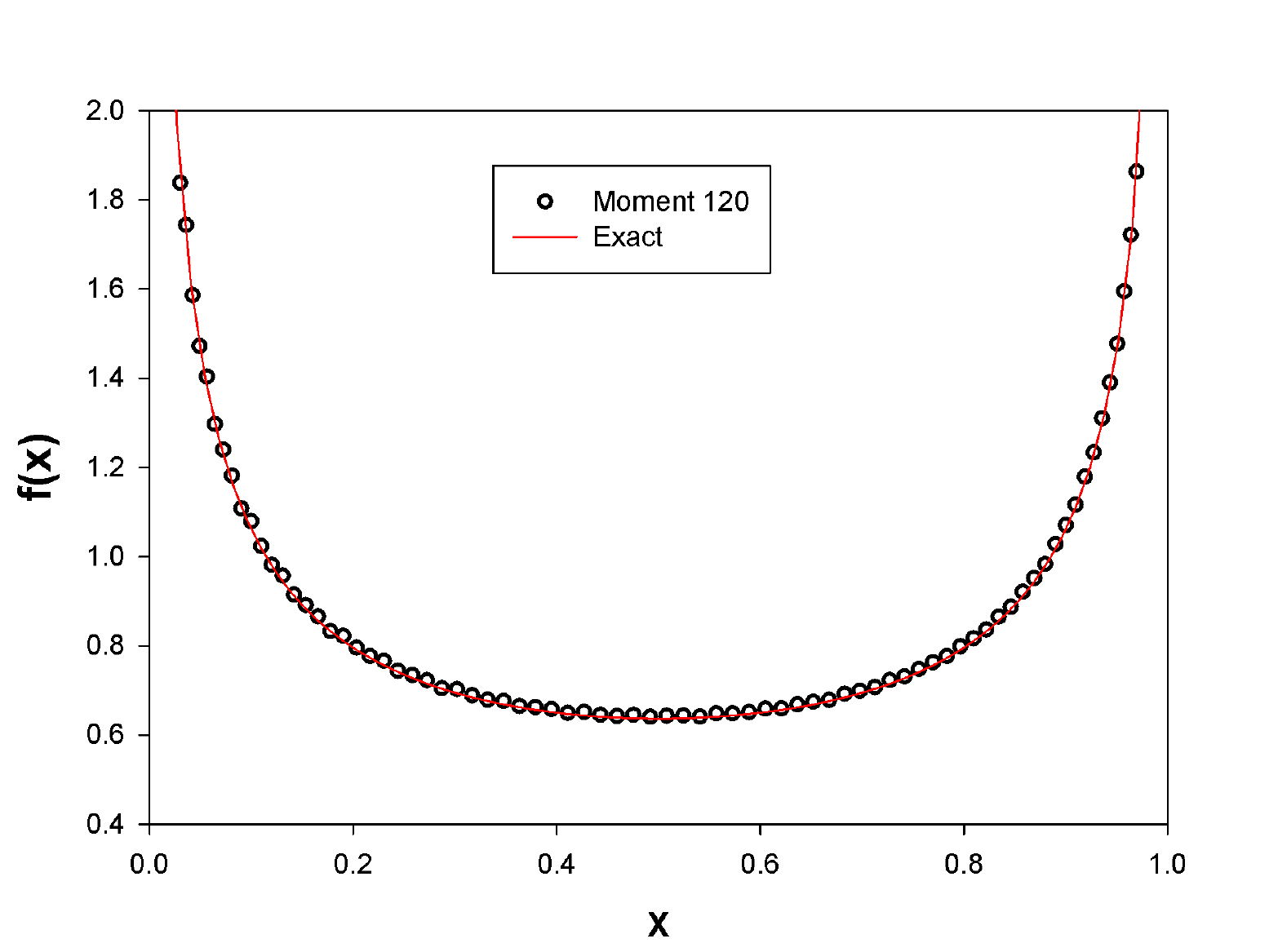}
\caption{
\label{fig9}
The reconstructed U-function using the first 120 moments along with the exact function 
evaluated at the quadrature points. The reconstructed function matches point-wise to 
the functional values as indicated in the figure. 
}
\end{figure}

\begin{figure}[htpb]
\includegraphics[width=4.5in, height=3.8in, angle = 0]{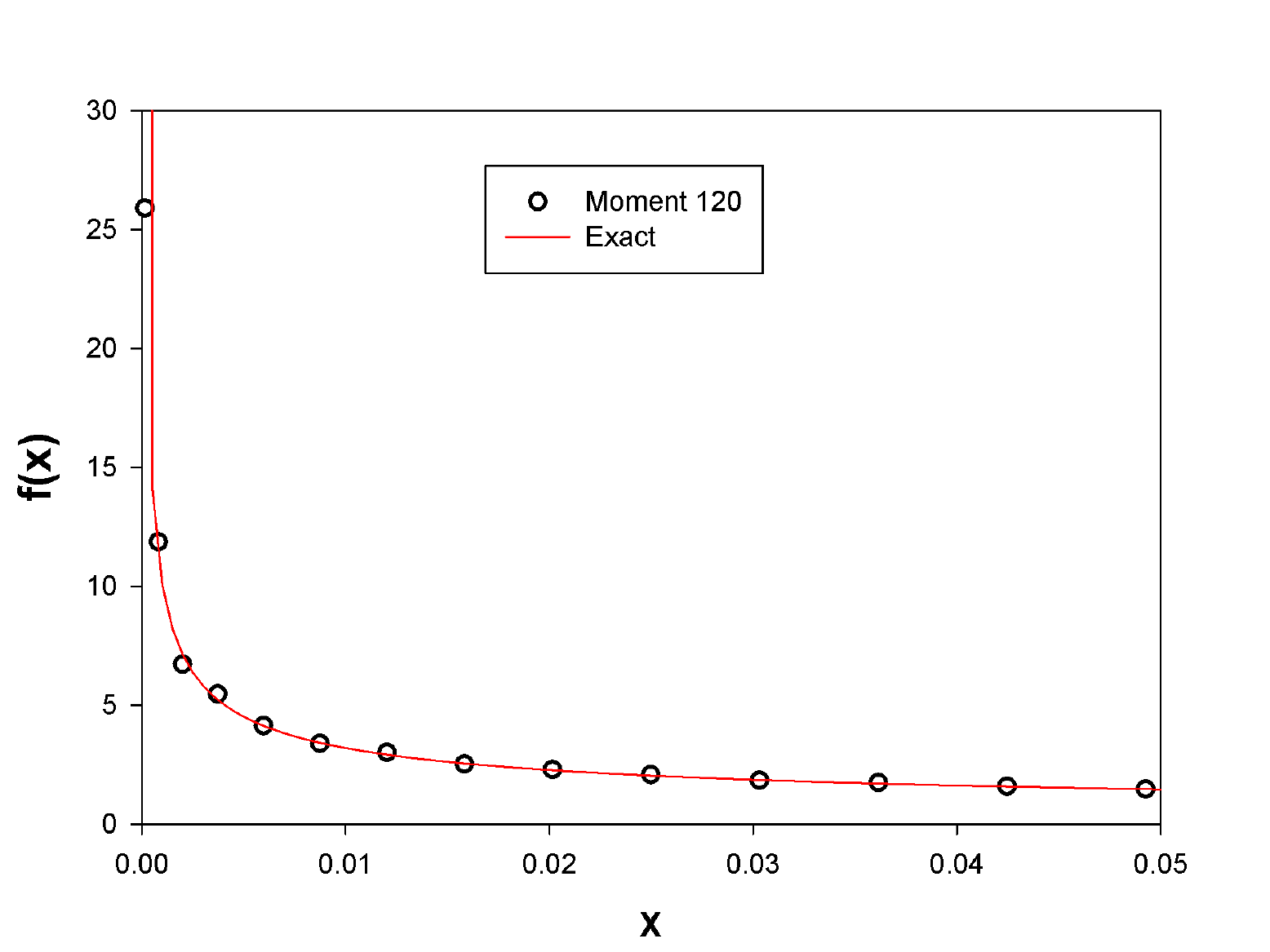}
\caption{
\label{fig10}
The divergent behavior of the U-function near $x=0$. The method 
accurately reproduces the function for very small values of $x$. 
}
\end{figure}

We now consider a function that has singularities in the range [0,1]. For the purpose of our 
discussion we refer to this function as `U-function' hereafter. The shifted Chebyshev moments 
of the function have the interesting property that except for the zeroth moment, all the other 
moments are identically zero.  The task of the ME algorithm in this case is to construct 
a function having all the moments zero except for the zeroth moment, which is unity by normalization. 
It may be noted that the electronic density of states per atom $D(E)$  of an infinite chain with a 
nearest neighbor interaction can be expressed in the form, 

\[ 
D(E) = \frac{1}{\pi} \frac{1}{\sqrt{4\beta^2 - (E - \alpha)^2}},
\] 

where $\alpha$ and $\beta$ are the on-site and the nearest neighbor hopping integrals respectively.  The 
zeroth moment is unity, which implies that there is only one state associated with each atom. For 
$\alpha = 0$ and $\beta = \frac{1}{2}$, the density of states can be mapped onto the U-function within 
the interval [0:1], and our algorithm can be applied to reconstruct the latter. An important characteristic of the 
density of states (or distribution function) is that it diverges at the band edges (or at the end points). 
Since all the Chebyshev moments are zero aside from the zeroth moment, it is important to see if the algorithm is capable of generating the density with 
the correct diverging behavior at the (band) edges.  In fig.\ref{fig8} we have plotted the results for 
the function for three different sets of moments $M$ = 10, 40, and 80 to illustrate how the approximate 
solutions improve with the increase of the number of moments. The shape of the function begins to emerge correctly even for as few as 
first 10 moments but with significant oscillations and poor divergence behavior near the end points. As the 
number of moments increases, the solution rapidly 
converges and the oscillations begin to disappear. In fig.\ref{fig9} we have plotted the results for 
$M = 120$.  The reconstructed function matches excellently throughout the interval with the exact one. 
The behavior of $f(x)$ near the left edge at $x=0$ is shown in fig.\ref{fig10} from $x$=0 to $x$=0.05. It is evident 
from the plot that even for very small values of $x$ near the left edge, the reconstructed values agree with 
the exact values excellently. A similar behavior has been observed near the right edge of the band near $x$=1. 
The capability of our method in reconstructing a function with singularities in the interval is 
thus convincingly demonstrated.

\subsection{Case 5: A function with a finite discontinuity}  
\begin{figure}[htpb]
\includegraphics[width=4.5in, height=3.8in, angle = 0]{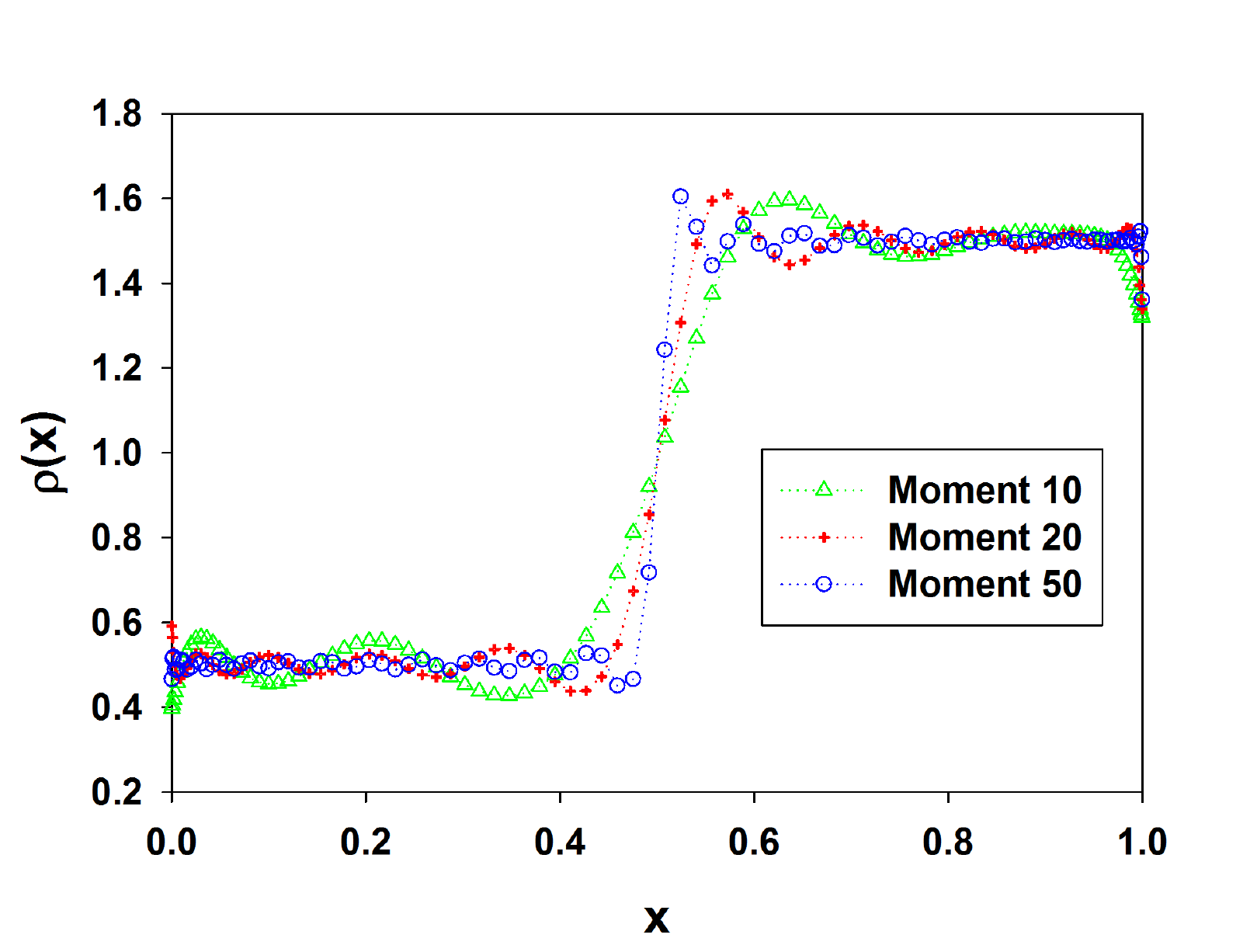}
\caption{ (Color online) The reconstructed double-step function for the first 10, 20 and 
50 moments. The reconstructed function improves progressively with the increase in the 
number of moments. 
\label{fig11}
} 
\end{figure}

\begin{figure}[htpb]
\includegraphics[width=4.5in, height=3.8in, angle = 0]{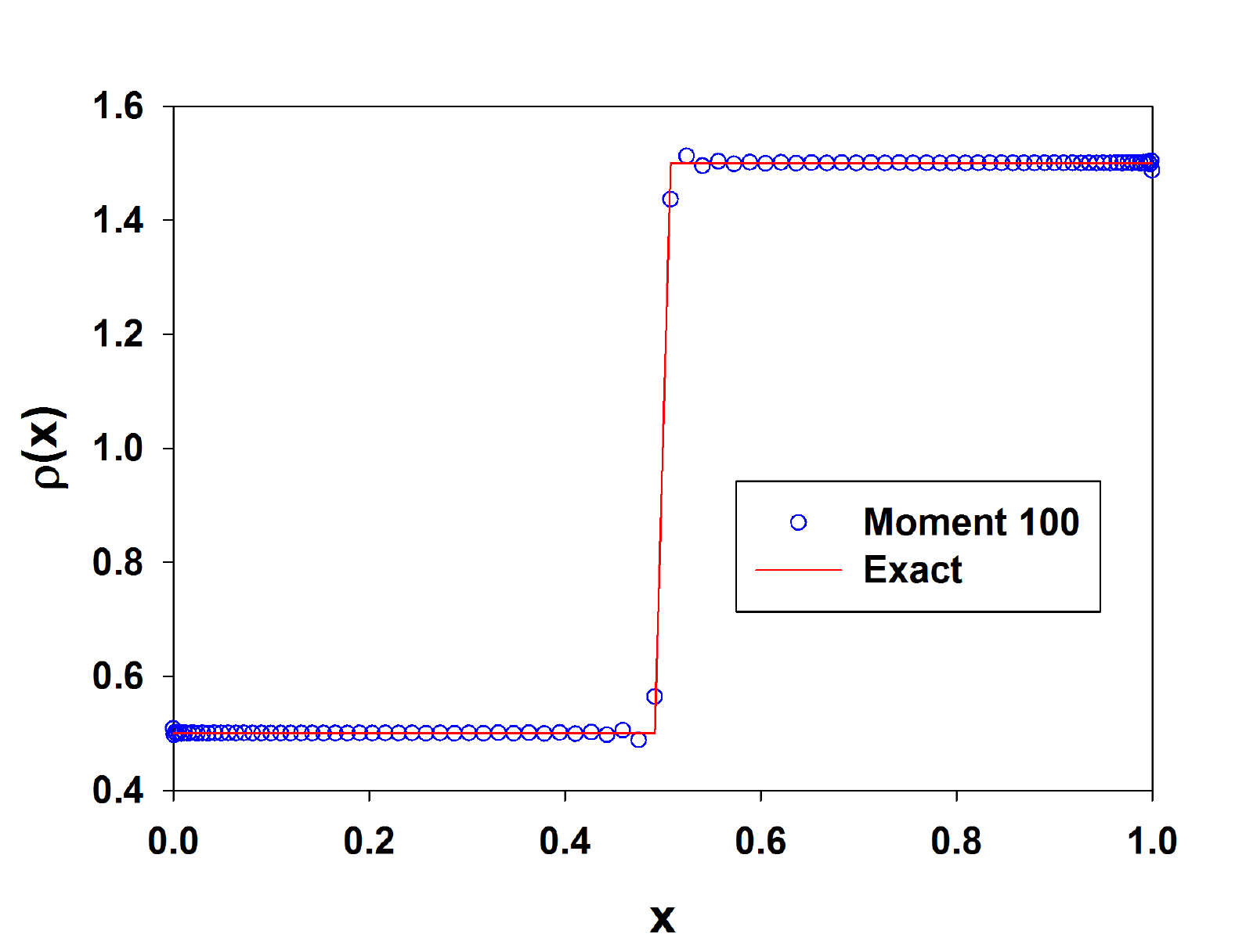}
\caption{ 
(Color online) The reconstructed double-step function using the first 100 moments. The exact 
function is also shown as a line. 
\label{fig12}
}
\end{figure}

The functions that we have discussed so far in the examples above are continuous within the 
interval. It would be interesting to consider a case where the function has a finite 
discontinuity within the interval. As an example, we choose a double-step function, 

\begin{equation*} 
f(x) = 
\begin{cases}   
\frac{1}{2} & \text{for $x  \le x_1$} \\ \\
\frac{3}{2} &  \text{for $x \ge x_1$,} 
\end{cases} 
\end{equation*} 

which has a finite discontinuity at $x_1 = \frac{1}{2}$. It is rather challenging to reconstruct 
the function from the moments so that the local behavior near the discontinuity at $x = 1/2$ is 
correctly reproduced. As before, the moment integrals can be calculated analytically in this case.  
In fig.\ref{fig11} we have plotted the function for 10, 20 and 50 moments. The solutions for the 
first two sets are expected to be less accurate, and indeed they show significant oscillations in the figure.  
For 50 moments the match is quite impressive.  On adding further moments, the solution  progressively improves.  
Figure \ref{fig12} shows the remarkable accuracy with which the function is reproduced by employing the first 100 moments. 
An important feature of the reconstructed function is that the discontinuity has been correctly 
reproduced with the exception of two points. From the various cases studied so far, we conclude 
that about 80 to 120 moments are needed for point-wise matching of the exact and the reconstructed 
functions.

\subsection{Case 6 : An unknown density }
\begin{figure}[htpb]
\includegraphics[width=4.5in, height=3.8in, angle = 0]{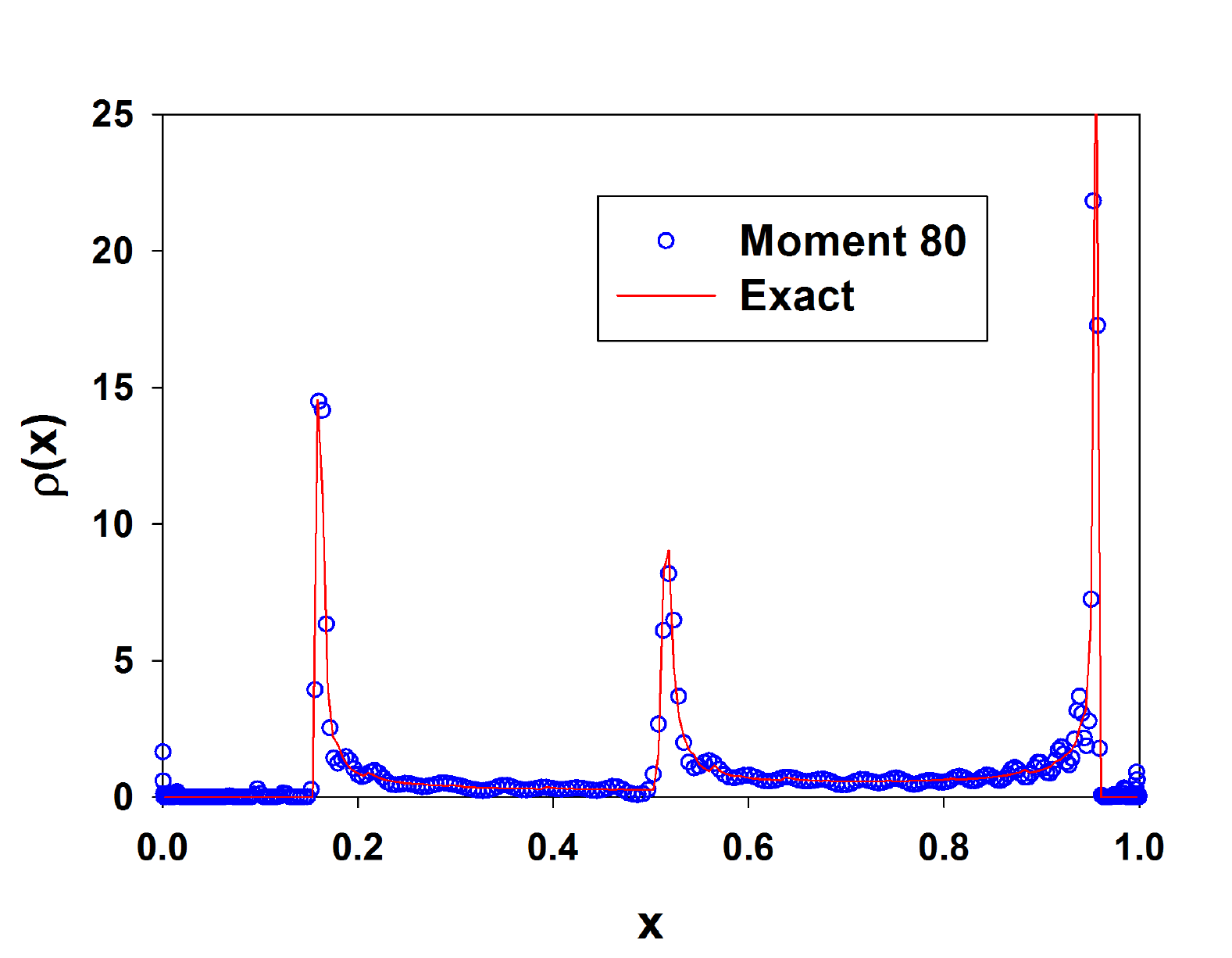}
\caption{A reconstructed density with sharp peaks obtained from the first 80 moments. 
The distribution corresponds to the natural invariant density of the logistic map as 
discussed in the text. The line corresponds to the numerical density obtained via 
histogram method. 
\label{fig13}
}
\end{figure}

Up until now, we have considered cases where the exact form of the function is known. In practical problems, 
however, it is more likely that the exact function is not available. We should therefore consider 
a case where the analytical expression for the distribution is not known, but a direct numerical solution is 
possible. As an example of such a distribution, 
we choose the natural invariant density of the logistic map $g(x) = \Gamma \, x \, (1-x)$ with $\Gamma=3.6785$. 
The invariant density for the map can be obtained by calculating the moments from the time 
evolution of an ensemble of initial iterates $x_0$ as discussed in Ref.\,\cite{Beck}. Since the map 
is ergodic for this value of $\Gamma$, the moments obtained via the time evolution of the map are identical 
to the moments of the natural invariant density~\cite{Beck, Steeb}. The task of our maximum entropy algorithm 
is to reconstruct the approximate density, and to compare it with the numerical density. The latter can be 
obtained from a histogram of the iterates and averaging over a large number of configurations~\cite{Beck}. 
The result from our ME ansatz using the first 80 moments is plotted in fig.\ref{fig13} along with the numerical 
density. The plot clearly demonstrates that every aspect of the fine structure of the numerical density is 
reproduced excellently in the maximum entropy solution. 

\begin{figure}[htpb]
\includegraphics[width=4.5in, height=3.8in, angle = 0]{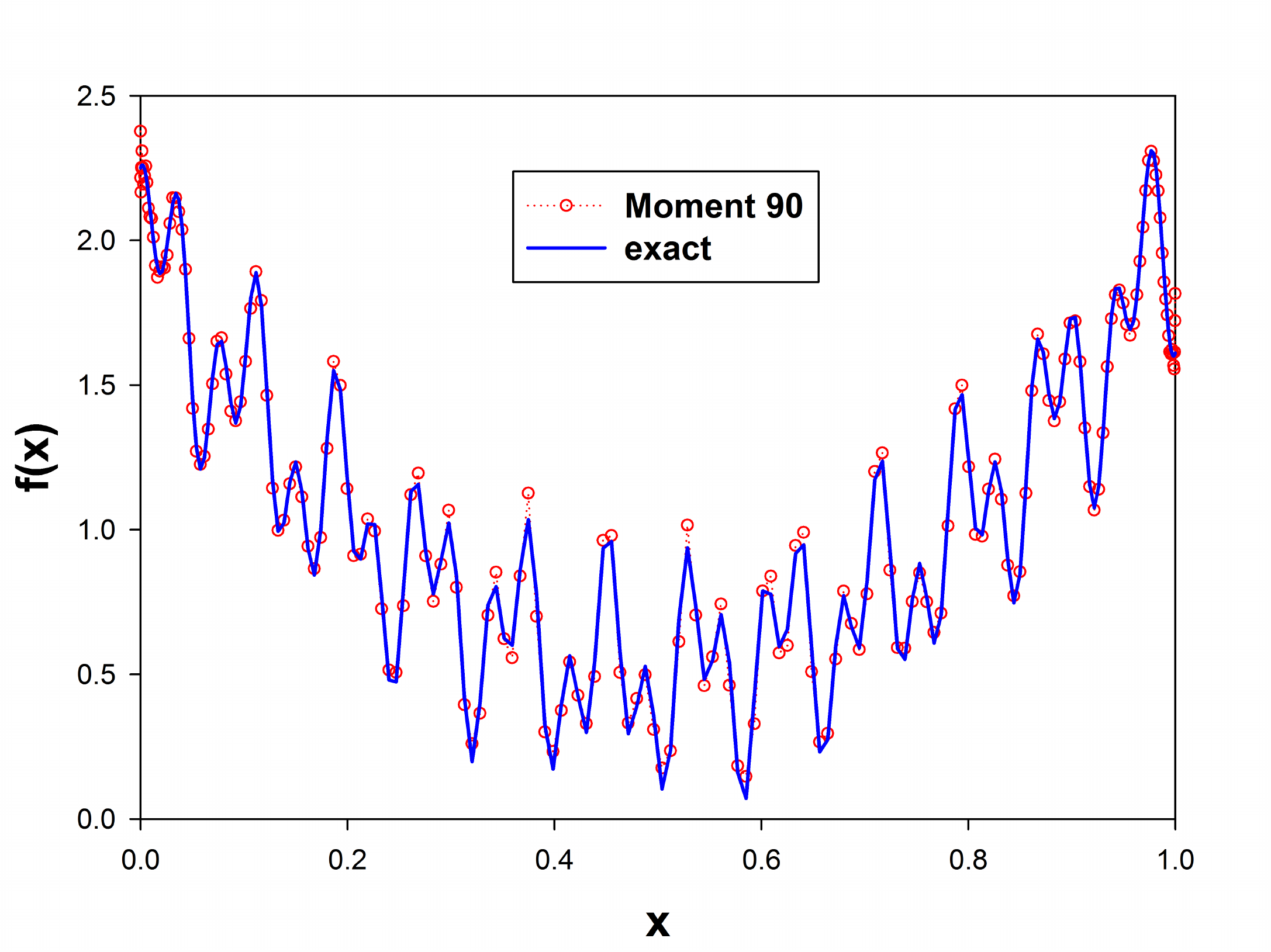}
\caption{
\label{fig14}
(Color online) 
The reconstruction of an oscillatory function with a fine structure as discussed in the text. 
The exact functional values are also plotted at the quadrature points for comparison. The 
location of the local minima and maxima are excellently reproduced from the first 90 moments 
of the function. 
}
\end{figure}

Finally, we end this section considering a rapidly oscillatory function having complex structure 
within the interval [0,1].  An example of such a function can be constructed as, 
\be
f(x) = \frac{1}{4} (\sin(167 x) + \cos(73 x)) + 6\,(x - \frac{1}{2})^2 + \frac{1}{2}
\label{oscill}
\ee
where the prefactors are chosen to normalize the function. In the context of studying diffusion in 
a rough one-dimensional potential, Zwanzig has studied such a function to obtain a general expression for the effective 
diffusion coefficient by analyzing the mean first-passage time~\cite{Zwanzig}. 
The maximum entropy construction of this function is plotted in fig.\ref{fig14} for 90 moments 
along with the exact function. Once again the function is reproduced excellently with every little 
details of the local minima and maxima of the function.

\section{Convergence behavior of the reconstructed solution} 
The convergence behavior of the maximum entropy solution has been discussed at length in 
the literature~\cite{Kullback1, Toussaint, Tag, Borwein}. The analytical  efforts are particularly 
focused on constructing bounds of the proximity of the reconstructed density to the exact 
density assuming that a given number of moments of the distributions are identical. 
In particular, given a target distribution $f(x)$ and a reconstructed distribution $f_M(x)$ that 
have identical first $M$ moments $\mu_0 = 1, \mu_1, \ldots, \mu_M$, the proximity of the two 
distributions can be expressed via the Kullback-Leibler divergence~\cite{Kullback1} and 
the variation measure~\cite{Tag}, 

\begin{subequations}\label{mea}
\begin{align}
D_{KL}[f, f_M] &=\int_{S} \, f(x) \ln \left(\frac{f(x)}{f_M(x)}\right)\, dx \label{first} \\
D_{v}[f, f_M] &= \int_{S} \, \vert f_M(x) - f(x) \vert \, dx \label{second}, 
\end{align}
\end{subequations} 
\noindent 
where $S$ is the support of the densities $f(x)$ and $f_M(x)$. The divergence measure is also 
known as the relative entropy or information discrimination, and $D_{KL} \ge 0$ with the equality 
holding if and only if  $f(x) = f_M(x)$ for all $x$. A lower bound for the divergence measure $D_{KL}$ in terms of the 
variation measure was given by Kullback~\cite{Kullback2}: 
\be 
D_{KL} \ge  \frac{D_v^2}{2} + \frac{D_v^4}{12}, 
\label{kl-tag}
\ee 
\noindent 
where it was assumed that the first $M$ moments are identical for both the distributions.  
Since the exact distributions are known for the examples considered here (except for the case 6), 
we can use these measures to examine if the reconstructed solution indeed satisfies the 
inequality. To this end, we first study the convergence of the moments of the reconstructed distributions 
with iteration and establish that the moments can be matched very accurately so that for practical purposes 
the reconstructed moments can be taken as identical to the exact (input) moments for the calculation of 
measure in Eqs.\,(\ref{first}) and (\ref{second}). 

\subsection{Convergence with respect to the number of iteration}
\begin{figure}[htpb]
\includegraphics[width=4.5in, height=3.8in, angle = 0]{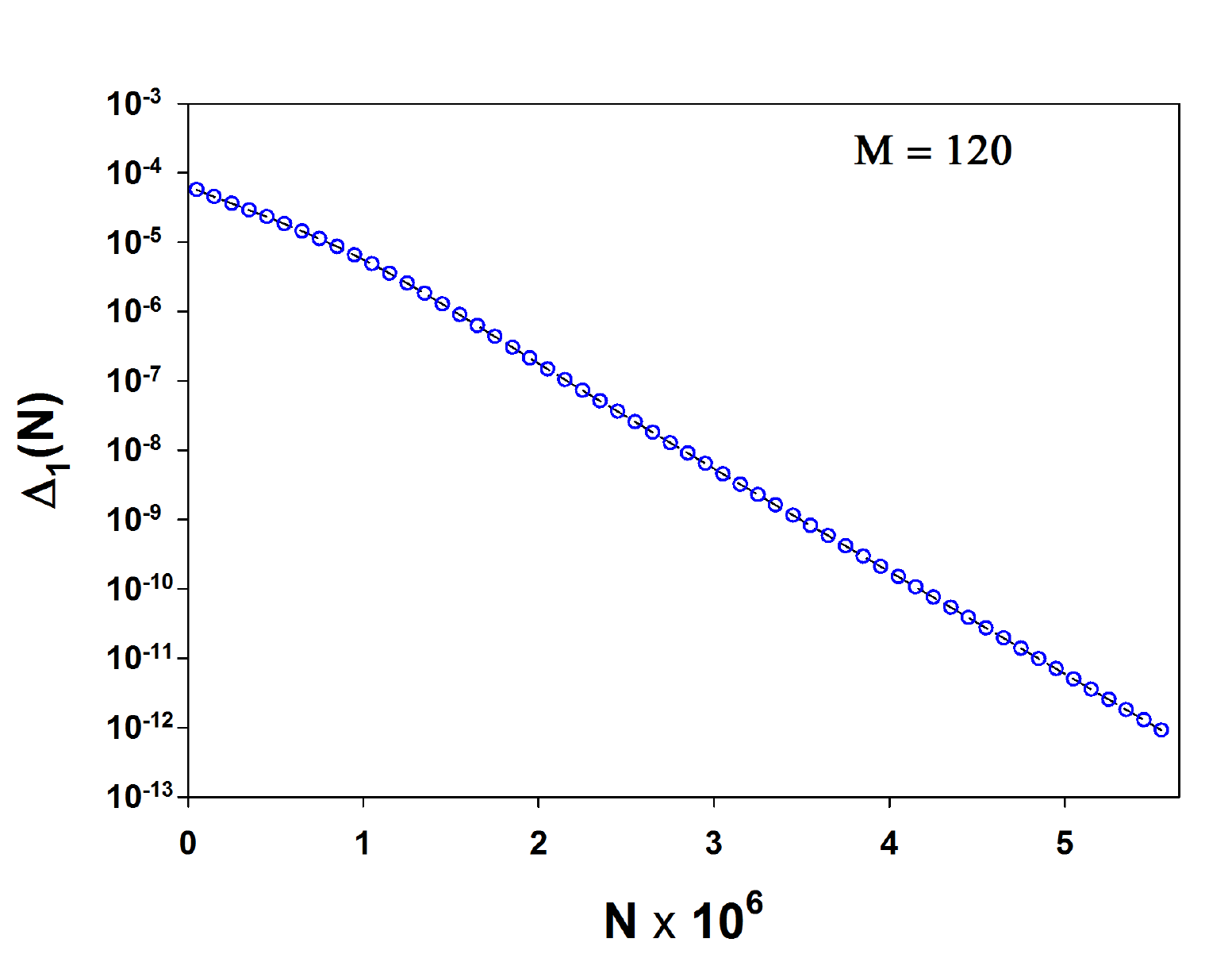}
\caption{
\label{fig15}
The semi-log plot of the RMS deviation $\Delta_1 (N)$ for the U-function with iteration $N$ expressed in unit of 
$10^6$. The RMS values decay exponentially with iteration after an initial crossover around 
$N = 0.8$. For clarity of presentation, every second data point is plotted in the figure. The 
number of moments is indicated in the figure. 
}
\end{figure}

As mentioned earlier, we have observed that the quality of the ME solutions depend on two 
factors: 1) the number of iterations and 2) the number of moments used for the purpose of reconstruction. 
In general, for a distribution with a fine structure, it is difficult to determine the minimal number of moments 
that are needed to reconstruct the function accurately. However, by studying a number of distributions 
with varying complexities and their convergence behavior, it is possible to obtain some useful information 
about the rate of convergence. We address this issue by choosing the U-function (case 4) as an example, but 
the observation is found to be true for other cases as well. 
For a systematic study of convergence behavior of the reconstructed 
moments with iterations, one requires a measure of the goodness of the fit. We therefore introduce $\Delta_1$, 
the root mean square (RMS) deviation of the exact moments from the moments provided by the ME solution, 
\be
\Delta_1(N, M) = \sqrt{\frac{1}{M} \sum_{i=1}^{M} (\mu_i - \tilde \mu_i(N, M))^2 }. 
\ee 
Here $\mu$ and $\tilde \mu(N, M)$ respectively denote the exact (or input) and the reconstructed (or output) 
moments, and the latter depends on the number of moments ($M$) and the iteration number ($N$). In the context of our present 
study, the exact moments of the functions are known, but in many practical cases they may not be available and 
need to be replaced by the input moments available for the problem.  The quantity $\Delta_1$ provides a measure of the proximity of 
the first $M$ reconstructed moments to the exact ones, and a small value of $\Delta_1$ is indicative of the 
fact that the moment constraints are satisfied with a high degree of accuracy. 
The value of $\Delta_1$ becomes as small as $10^{-14}$ provided an adequate number of iterations are performed to match a given set 
of moments. 
In fig.\ref{fig15}, we have plotted $\Delta_1$ for the case of U-function with the number of iteration progressively 
increasing to $N = 6 \times 10^{6}$.  The RMS deviation decreases rapidly with the number of iteration and eventually drops to a value of the order 
of $10^{-14}$.  An examination of the data suggests that $\Delta_1$ can be fitted to an exponential decay with 
iteration and is plotted in fig.\ref{fig15}. This behavior is also observed for the other distributions 
discussed in Section 3. It thus follows that the algorithm converges quite rapidly, and that the moment constraints 
can be satisfied to a high degree of accuracy even for a very large moment set. 

\subsection{Convergence with respect to the number of moments}
\begin{figure}[htpb]
\includegraphics[width=4.5in, height=3.8in, angle = 0]{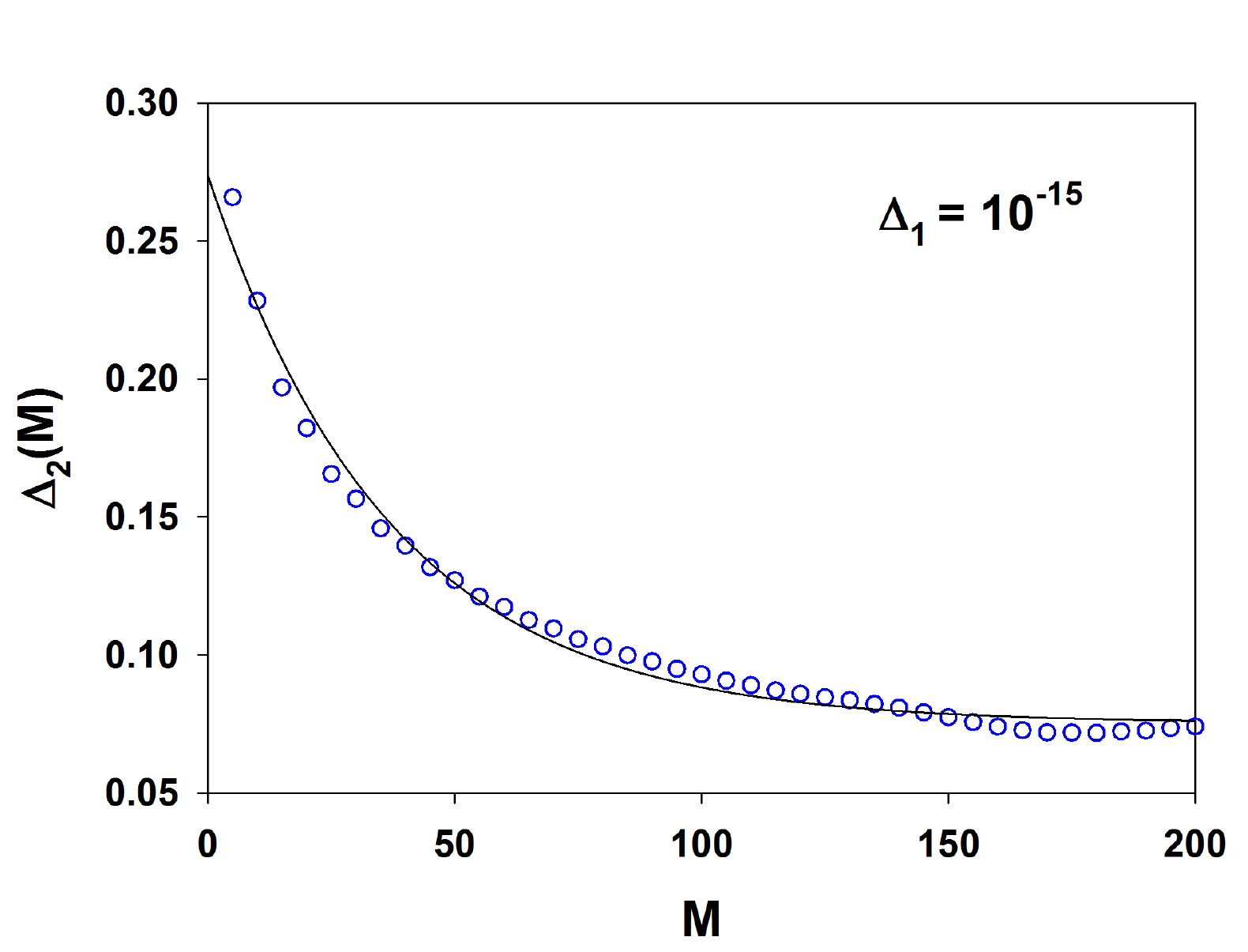}
\caption{
\label{fig16}
The variation of the RMS deviation $\Delta_2(M)$ with moments for a given value of 
$\Delta_1 = 10^{-15}$. The data can be fitted to exponential decay as indicated by 
the best fitted line in the plot. 
}
\end{figure}

While $\Delta_1$ provides a measure of the goodness of the fit for the moments, it does not necessarily 
guarantee a point-wise convergence of the reconstructed function with the exact one. This is 
particularly true if a small number of moments are used to reconstruct the function that has a fine structure 
in it (cf. fig.\ref{fig13}).  In this case, the reconstructed moments can be matched to a high degree of 
accuracy with input moments, but the solution may still miss out the characteristic feature of 
the distribution folded in the higher order moments. The maximum entropy solution in this case may not reproduce 
the actual solution even though the approximate moments are very close to the exact moments. To ensure that 
$\Delta_1$ indeed attains a sufficiently small value, we need to study the approximate solution vis-a-vis 
the number of moments for a fixed cycle of iterations. Since the exact functions are known in our cases, 
the simplest way to measure the quality of the ME solution is to construct the RMS deviations of the 
reconstructed functions from the exact ones in the interval [0,1]: 
\bea
\Delta_2(N, M) &=& \sqrt{\frac{1}{n_g} \sum_{i=1}^{n_g} \left[f_i - \tilde f_i (N, M)\right]^2} \\
               &\approx & \Delta_2(M) \quad \mbox{\rm for large N,} \nonumber 
\eea 
where $n_g$ is the number of points used in the quadrature. Here we have assumed that 
the dependence of $\Delta_2$ on $N$ can be neglected so that $\tilde f_i(N, M) \approx \tilde f_i(M)$, 
which holds for large $N$ owing to the fast decay of $\Delta_1$.  We choose $\Delta_1 = 10^{-15}$ for 
each of the moment sets to study the variation of $\Delta_2(M)$ for different values of $M$. 
In practice, the exact function may not be available 
but the expression for $\Delta_2$ can still be used by replacing the exact function $f_i$ by 
$\tilde f_i(M + \Delta M)$ and constructing the RMS deviation for increasing values $M$ and 
$\Delta M$. In fig.\ref{fig16}, we have plotted $\Delta_2$ for the case of U-function for different values of M. 
The plot shows a monotonic 
decrease of $\Delta_2$ with the increasing values of M.  For this function, we see that a value of 
$M$=100 to 120 provides a small enough $\Delta_2$ to reconstruct the function accurately when 
$\Delta_1 = 10^{-15}$. The solid line in the figure is an exponential fit to the data indicating a 
fast convergence of our algorithm with respect to the moments for a fixed value 
of $\Delta_1$. 

\begin{figure}[htpb]
\includegraphics[width=4.5in, height=3.8in, angle = 0]{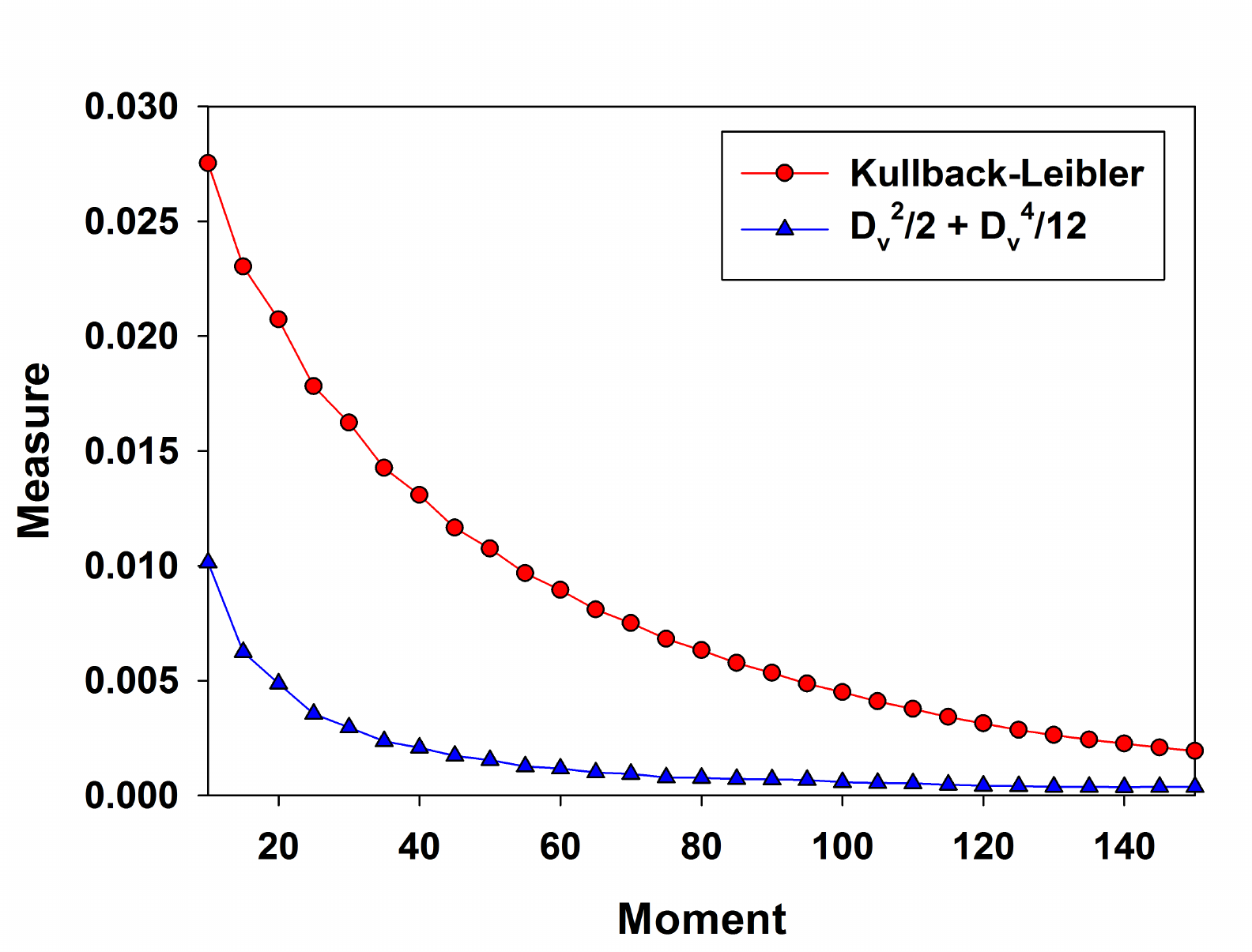}
\caption{
\label{fig17}
The variation of the KL divergence measure (circles) of the U-function for different 
number of moments.  The function of the variation measure from the right hand side of the inequality 
(\ref{kl-tag}) is also plotted for comparison. The data correspond to the reconstructed solution 
with $\Delta_1 = 10^{-15}$. 
}
\end{figure}

The proximity of the reconstructed distribution to the exact one can be quantified in terms of 
the divergence measure and the variation distance as defined in the beginning of this section. A 
number of inequalities can be found in the literature~\cite{Tag, Toussaint} that provide lower 
bounds of the relative entropy. The inequality in (\ref{kl-tag}) is an example of such a bound 
although still sharper bounds are available in the literature~\cite{Tag, Toussaint}. 
In fig.\ref{fig17} we have plotted the relative entropy of the U-function for different number of moments. The reconstructed 
solution for each of the moment sets M corresponds to $\Delta_1 = 10^{-15}$ so that the first M moments of 
the exact and the reconstructed functions are practically identical to each other. The right hand side of the inequality 
(\ref{kl-tag}) is also plotted in the same figure for comparison. As the reconstructed solution approaches 
the exact solution with increasing number of moments, the relative entropy or information discrimination 
between the two distributions decreases and eventually comes very close to the analytically predicted 
lower bound.

\section{Conclusion} 
In this paper we study the reconstruction of functions from a set of Chebyshev moments (of the 
functions) via maximum entropy optimization. The method consists of mapping the original 
constraint optimization problem in primal space onto an unconstrained convex optimization 
problem in the dual space by using a discretized form of the moments and the Shannon entropy 
of the function to be reconstructed. The resulting optimization problem is then solved iteratively 
by obtaining the optimal set of Lagrange's parameters as prescribed in Ref.\cite{Biswas}. 
By virtue of its ability to deal with a larger number of moments, our present approach is extremely 
robust and accurate. This makes it possible to reconstruct a variety of function that are difficult to handle 
otherwise. 

We demonstrate the accuracy of this method by applying to a number of functions for which the exact 
moments are available. The method accurately reproduces not only smooth and continuous functions 
(such as square-root and double-parabolic functions) but also non-smooth and discontinuous 
functions (such as a double-step function with a finite discontinuity).  It also captures 
the fine structure in a rapidly oscillatory function of known analytical form and 
the invariant densities of a logistic map corresponding to special values of the control 
parameter for which no analytical results are available. 
A convergence study of the reconstructed moments suggests that the RMS deviation of the moments (from 
the exact ones) can be made as small as $10^{-15}$ indicating the accuracy with which the input 
moments can be matched with the reconstructed ones.  
A direct comparison with the exact functions studied here reveals that the method indeed converges 
to the correct solution provided a sufficient number of moments are available as input. 
The general trend of the convergence profile is similar in all the cases:  the quality of the reconstructed function markedly 
improves with the increase in the number of moments. The numerical calculations suggest that the convergence 
toward the exact solution is almost of exponential nature.

\acknowledgments
The work is partially supported by a fellowship from Aubrey Keith Lucas and Ella Ginn Lucas 
Endowment at the University of Southern Mississippi.  PB would like to thank Lawrence Mead, 
David Drabold, and Roger Haydock for useful discussions during the course of the work.


\begin{thebibliography}{99}

\bibitem{Shohat}
A. J. Shohat and J. D. Tamarkin, {\it The Problem of Moments} (American Mathematical Society, 1963)

\bibitem{Akheizer}
N. I. Akheizer,{\it The classical moment problem and some related questions in analysis}
(Hafner Publishing Co., New York, 1963)

\bibitem{Kapur} 
J. N. Kapur and H. K. Kesavan, {\it Entropy optimization and mathematical programming}, 
Kluwer Academic Publishers, Dordrecht (1997)

\bibitem{Lent}
A. Lent in {\it Image analysis and evaluation} (Edited by R. Shaw), SPSE, Washington, D.C (1953)

\bibitem{Wheeler}  
J. C. Wheeler and R.G. Gordon, J. Chem. Phys. {\bf 51}, 5566 (1969) 

\bibitem{Isenberg}
C. Isenberg, Phys. Rev. {\bf 150}, 712 (1966) 

\bibitem{Bricogne} 
G. Bricogne, Acta Cryst. A {\bf 44}, 517 (1988) 

\bibitem{Gilmore} 
C. J. Gilmore, Acta Cryst. A {\bf 52}, 561 (1996) 

\bibitem{Mead}
L.R. Mead and N. Papanicolaou, J. Math. Phys. {\bf 25}, 2404 (1984) 

\bibitem{Smith}  
C.R. Smith and W.T. Grandy Jr., {\it Maximum entropy and Bayesian methods
in inverse problems}, Reidel, Dordrecht (1985)

\bibitem{Drabold} 
D.~A.~Drabold and O.~F.~Sankey, Phys.~Rev.~Lett. {\bf{70}}, 3631 (1993)

\bibitem{Carlsson} 
A.E. Carlsson and P.A.Fedders, Phy. Rev. B {\bf 34}, 3567 (1986) 

\bibitem{Gotovac} 
H. Gotovac and B. Gotovac, J. Comp. Phys. {\bf 228}, 9079 (2009)

\bibitem{Hausdorff} 
F. Hausdorff, Math. Z. {\bf 16}, 220 (1923) 

\bibitem{Hadamard} 
J. Hadamard, {\it Lectures on the Cauchy problem in linear partial differential 
equations}, Yale University Press, New Haven  1923 

\bibitem{Turek} 
I. Turek, J. Phys. C: Solid State Phys. {\bf 21}, 3251 (1988) 

\bibitem{Tikhonov} 
A. Tikhonov and V. Y. Arsenine, {\it Solution of ill-posed problems}, V.H. Winston \& Sons, Washington, D. C. (1977) 

\bibitem{Vino1} 
G.A. Viano, J. Math. Anal. Appl. {\bf 156}, 410 (1991)

\bibitem{Vino2} 
E. Scalas and G. A. Viano, J. Math. Phys. {\bf 34}, 5781 (1993) 

\bibitem{Jaynes}
E.T. Jaynes, Phys. Rev. {\bf 106}, 620 (1957) 

\bibitem{Shannon}
C. Shannon, Bell Syst. Tech. J. {\bf27}, 379 (1948)

\bibitem{Schoenfeldt} 
J-H. Sch\"ofeldt, N. Jimenez, A.R. Plastino, A. Plastino and M. Casas, Physica A {\bf 374}, 573 (2007) 

\bibitem{Steeb} 
W-H. Steeb, F. Solms and R. Stoop, J. Phys. A: Maths. Gen {\bf 27}, L399 (1994) 

\bibitem{Maradudin} 
A.A.Maradudin, E.W.Montroll, G.H.Weiss, and I.P.Ipatova, {\it Theory of lattice dynamics in harmonic approximation} 
(Academic Press, New York, 1971) 

\bibitem{Houston} 
W.V. Houston, Rev. Mod. Phys. {\bf 20}, 161 (1948) 

\bibitem{Domb} 
C. Domb and C. Isenberg, Proc. Phys. Soc. London {\bf 79}, 659 (1962) 


\bibitem{Silver} 
R.N. Silver and H. R\"{o}der, Phys. Rev. E {\bf 56}, 4822 (1997)


\bibitem{Biswas}
K. Bandyopadhyay, A.K. Bhattacharya, P. Biswas and D.A. Drabold, Phys. Rev. E {\bf71}, 057701 (2005)

\bibitem{Kullback1} 
S. Kullback, {\it Information theory and statistics} (Dover Publication, 1997) 

\bibitem{Tag} 
A. Taglinai, Appl. Math. and Comput. {\bf 145}, 195 (2003) 

\bibitem{Wimp} 
J. Wimp, Proc. Roy. Soc. Edinburgh {\bf 82 A}, 273 (1989)

\bibitem{note1} 
The solution is {\it least biased} as far as the entropy of the density function is 
concerned, and the Hausdorff conditions are satisfied. There is no guarantee that 
the maximum entropy solution would be close to the exact solution, particularly when 
the very first few moments are used in the reconstruction procedure. 

\bibitem{Berg}
L.~M.~Bergman, USSR Comput.~Math. and Math. ~Phys. {\bf 7}, 200 (1967)

\bibitem{note2} 
The exact location of the abscissa is computed by linearly interpolating between the two 
ordinates that have zero and non-zero values. The accuracy (of the location of the abscissa) 
can be improved further by using more points in the quadrature formula. 

\bibitem{Beck}
C. Beck and F. Schl\"{o}gl, {\it Thermodynamics of chaotic systems} (Cambridge University 
Press, Cambridge, United Kingdom, 1993)

\bibitem{Zwanzig} 
R. Zwanzig, Proc. Nat. Acad. Sci. USA {\bf 85}, 2029 (1988) 

\bibitem{Borwein} 
J.M. Borwein, SIAM J. Optim {\bf 1}, 191 (1991) 


\bibitem{Toussaint} 
G. T. Toussaint, IEEE Trans. Inform. Theor. {\bf 21}, 99 (1975) 

\bibitem{Kullback2} 
S. Kullback, IEEE Trans. Inform. Theor. IT-13, 126 (1967) 



\end{thebibliography}
\end{document}